\documentclass{aa} 

\usepackage{booktabs}
\usepackage{threeparttable}
\usepackage{graphicx}
\usepackage{txfonts}
\usepackage{bm}
\usepackage{ulem}
\usepackage{makecell}
\usepackage{hyperref}
\hypersetup{
    colorlinks=true,
    citecolor=blue,
    linkcolor=red,
    filecolor=magenta,
    urlcolor=cyan,
}

\makeatletter
\renewcommand*\aa@pageof{, page \thepage{} of \pageref*{LastPage}}
\makeatother
\usepackage{amstext}

\begin{document}

\title{Benchmark tests of transmission spectroscopy using transiting white dwarfs}

\author{C. Jiang \inst{1,2} \fnmsep\thanks{\email{czjiang@pmo.ac.cn}}
        \and
        G. Chen \inst{1,3} \fnmsep\thanks{\email{guochen@pmo.ac.cn}}
        \and
        E. Pall\'{e} \inst{4,5}
        \and
        H. Parviainen\inst{4,5}
        \and
        F. Murgas\inst{4,5}
        \and
        Y. Ma \inst{1}
        }

\institute{
CAS Key Laboratory of Planetary Sciences, Purple Mountain Observatory, Chinese Academy of Sciences, Nanjing 210023, PR China
\and
School of Astronomy and Space Science, University of Science and Technology of China, Hefei 230026, PR China
\and
CAS Center for Excellence in Comparative Planetology, Hefei 230026, PR China
\and
Instituto de Astrof\'{i}sica de Canarias, V\'{i}a L\'{a}ctea s/n, E-38205 La Laguna, Tenerife, Spain
\and
Departamento de Astrof\'{i}sica, Universidad de La Laguna, Spain
}

\date{Received 8 Dec 2021 / Accepted 9 May 2022}

\abstract
{Ground-based transit observations are affected by both telluric absorption and instrumental systematics, which can affect the final retrieved transmission spectrum of an exoplanet. To account for these effects, a better understanding of the impact of different data analyses is needed to improve the accuracy of the retrieved transmission spectra.}
{We propose validating ground-based low-resolution transmission spectroscopy using transiting white dwarfs. These targets are selected to have transit parameters comparable with typical transiting hot Jupiters but nondetectable transmission signals due to their extremely high surface gravities. The advantage here is that we know beforehand what the final transmission spectrum should be: a featureless flat spectrum.}
{We analyzed two transiting white dwarfs analogous to hot Jupiters, KIC 10657664B and KIC 9164561B. We used various noise models to account for the systematic noise in their spectroscopic light curves following common procedures of transmission spectroscopy analyses. We compared the derived transmission spectra with the broadband transit depth to determine whether there are any artificial offsets or spectral features arising from light-curve fitting.}
{The results show a strong model dependence, and the transmission spectra exhibit considerable discrepancies when they are computed with different noise models, different reference stars, and different common-mode removal methods. Nonetheless, we can still derive relatively accurate transmission spectra based on a Bayesian model comparison.}
{With current ground-based instrumentation, the systematics in transit light curves can easily contaminate a transmission spectrum, introducing a general offset or some spurious spectral features and thus leading to a biased interpretation on the planetary atmosphere. Therefore, we suggest that any wiggle within the measurement errors in a transmission spectrum should be interpreted with caution. It is necessary to determine the dependence of results on the adopted noise model through model comparison. The model inferences should be examined through multiple observations and different instruments.}


\maketitle

\section{Introduction}
\label{sect_introduction}

    Transit spectroscopy is currently the most accessible and most widely used technique for revealing the absorption and scattering features of exoplanet atmospheres. Based on transmission spectra, we can identify the atomic and molecular species, along with the properties of clouds or hazes, in planetary atmospheres \citep{Seager2000, Brown2001, Madhusudhan2019}. When transit spectroscopy is carried out with ground-based telescopes, special attention must be paid to the systematics of telluric origins. For high-dispersion spectroscopy, the telluric absorption lines can be directly resolved and corrected \citep{Redfield2008, Snellen2008, Snellen2010}. For low-dispersion spectroscopy, the long-slit or the multi-object observational mode can be used to simultaneously monitor the fluxes of the target and the reference star(s), and then differential spectrophotometry can be applied to reduce the telluric effects. These methods have been widely employed in ground-based transit spectroscopic surveys \citep[e.g., ][]{Sing2012, Nikolov2016, Sedaghati2016, Chen2017b, Sedaghati2017, Chen2018, Nikolov2018, Espinoza2019, Kirk2019, Cater2020}
    
    We used spectroscopic light curves from ground-based observations to validate the methods commonly used in low-dispersion transmission spectroscopy. Our aim is to investigate to which extent the light-curve systematics would impact the derived transmission spectra. We performed this benchmark test by observing a white dwarf transiting a main-sequence star. The geometric features of transit light curves for white dwarfs are similar to those for hot Jupiters, except for the additional systematic effects from eclipsing binaries, including ellipsoidal light variation, mutual illumination, and Doppler boosting \citep{Carter2011}. According to \cite{Di2011}, the hydrogen envelop of the white dwarf progenitor is stripped off in the late phase of mass transfer with its main-sequence companion, leaving a degenerate helium core that is the white dwarf. Due to the extremely high surface gravities of white dwarfs, their atmospheric scale heights are so small that the transmission signals are undetectable in the visible to near-infrared wavebands, limited by current instrumental photometric precision, thus producing featureless transmission spectra.

    We selected two transiting white dwarfs as our targets of interest: KIC 10657664B (also known as KOI-964B), and KIC 9164561B. They both transit A-type dwarf stars, and their transit parameters are close to those of typical hot Jupiters. The binary system KIC 10657664 was selected for its well-determined transit parameters, where the white dwarf companion orbits the A-type host star ($2.23\pm0.12~\mathrm{M_\odot}$, $1.89\pm0.14~\mathrm{R_\odot}$, 13.64 R-mag) at a distance of $\text{about seven}$ times the host star radius, with a period of $\sim$3.27 days \citep{Carter2011, Wong2020}. Several reference stars were aligned, together with the target, inside the long slit of the spectrograph, which enabled us to investigate whether the selection of reference stars would impact the calculated transit parameters and transmission spectra. The other target, KIC 9164561B, orbits the primary star ($2.02\pm0.06~\mathrm{M_\odot}$, $2.54\pm0.03~\mathrm{R_\odot}$, 14.06 R-mag) at a closer distance of $\sim$2.5 times the host star radius with a shorter period of $\sim$1.27 days \citep{Rappaport2015}. This binary system was selected for its oscillating light curves that arise from the stellar pulsation of the A-type host, which can be treated as an analog to other transiting exoplanets whose host stars also exhibit pulsation features in the light curves \citep[e.g., WASP-33b;][]{Essen2014,Essen2019}. Based on this target, we investigated whether an unbiased transmission spectrum can be derived when the light curves are contaminated by considerable stellar pulsations.

    This paper is organized as follows. In the next section, we summarize the transit observations and the data reduction procedures. We introduce the methods of light-curve analyses in Sect. \ref{sect_method}. The results of benchmark tests are shown in Sect. \ref{sect_results}. We draw our conclusions in Sect. \ref{sect_conclusions}.

\section{Observations and data reduction} 
\label{sect_observations}

    The transit events of our two targets were observed using the long-slit mode of the Optical System for Imaging and low-Intermediate-Resolution Integrated Spectroscopy at the Gran Telescopio Canarias (GTC OSIRIS) with the R1000R grism. Two red optimized $2048 \times 4096$ Marconi CCDs were used to record images with a gap of $9.4''$ in between. The pixel scale of the imaging was $0.254''$ after a $2 \times 2$ pixel binning. The gap between the two CCDs was parallel to the dispersion direction and did not affect the OSIRIS spectra. The unvignetted field of view for imaging was $7.8' \times 7.8'$. The slit was $7'$ long and $12''$ wide. The total waveband of the R1000R grism is 5100 -- 10\,000 \text{\AA} with an instrumental resolution of 2.6 \text{\AA} per pixel. The standard 200 kHz readout mode was adopted, and the corresponding readout noise was $\sim$4.5 $\rm e^-$. The arc lines of HgAr, Xe, and Ne were measured through a $1''$ wide slit with the same R1000R grism for the wavelength calibration of the science spectra.    

\begin{figure*}
        \centering
        \includegraphics[width=\linewidth]{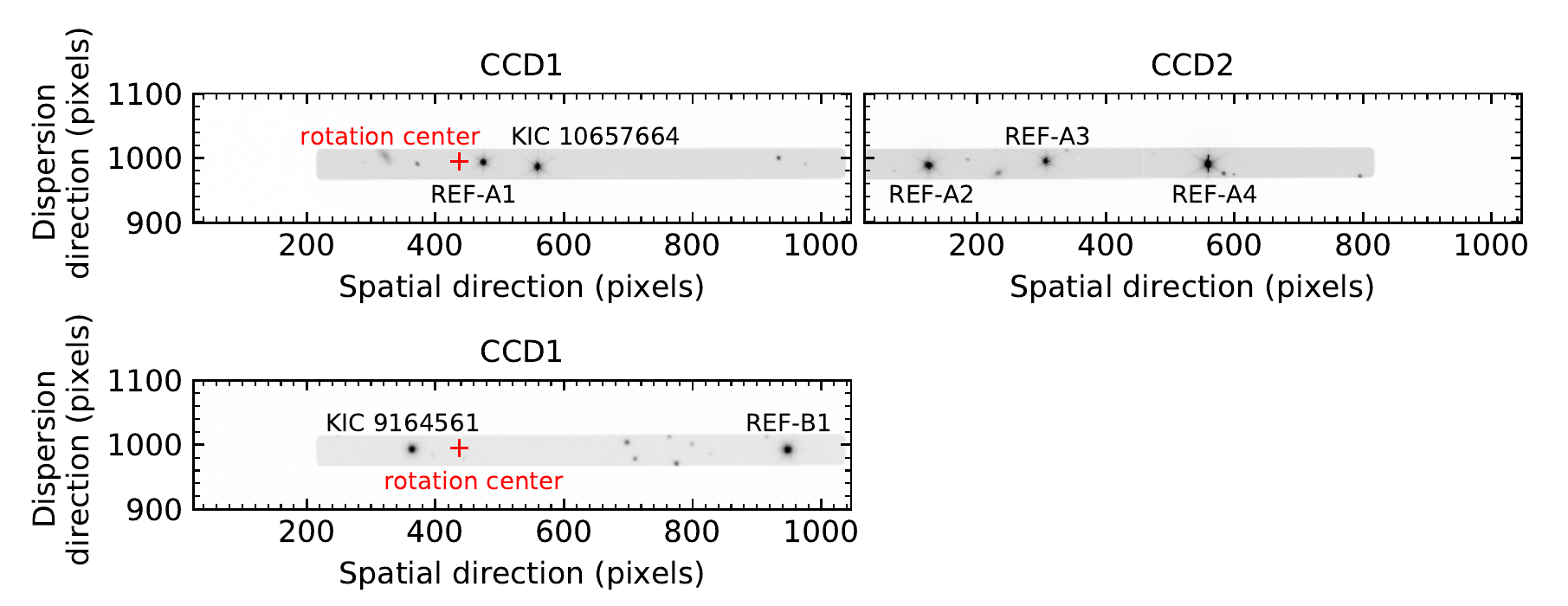}
        \caption{Through-slit images of the two observations. {\it Upper row}: Positions of KIC 10657664 and the reference stars (REF-A1 to A4) in CCD1 and CCD2. {\it Lower row}: Positions of KIC 9164561 and the reference star (REF-B1) in CCD1. The red mark is the GTC Nasmyth rotator center. Only a part of the data section is displayed for clarity.}
        \label{fig_slit}
    \end{figure*}

    The transit of KIC 10657664B was observed on 16 July 2015. During the observation, the Moon was absent and the weather was mostly clear. Four neighboring reference stars (hereafter REF-A1, REF-A2, REF-A3, and REF-A4) with similar magnitudes were observed simultaneously with the target star. Table \ref{table_stars} lists the identifiers, stellar types, and R-band magnitudes of all observed stars. The target and the REF-A1 star were placed on CCD1, while the other three stars (REF-A2, REF-A3, and REF-A4) were placed on CCD2 (Fig. \ref{fig_slit}). 
    
    The transit of KIC 9164561B was observed on 10 July 2015. The weather was clear during the observation, while the Moon was illuminated by $\sim$36\% at a distance of 87$^\circ$. One reference star KIC 9099798 (hereafter REF-B1) was simultaneously observed with the target star, and both of them were placed on CCD1 (see Fig. \ref{fig_slit}). Although REF-A3 and REF-B1 are rotationally variable stars, their variation periods \citep[$\sim$14.6 days for the former and $\sim$13.2 days for the latter;][]{McQuillan2014} are much longer than the duration of the observations, thus hardly affecting the differential flux correction. More details of the two observations are listed in Table \ref{table_observation}.
    
    \begin{table}[tbp]
        \renewcommand\arraystretch{1.2}
        \centering
        \begin{threeparttable}[b]
        \caption[]{Information of the target and the reference stars.}
        \label{table_stars}
        \begin{tabular}{lllc}
            \toprule
            Labels          & 
            KIC             & 
            Object types    & 
            R mag\tnote{(1)}  \\
            \midrule
            -           &
                10657664   &
                A1V (host star)\tnote{(2)}&
                13.64       \\
            REF-A1           & 
                10657673    & 
                G2V\tnote{(3)} &
                14.01       \\
            REF-A2           & 
                10723007    & 
                G4V\tnote{(3)}           &
                13.21       \\
            REF-A3           & 
                10722988    & 
                rotationally variable G1\tnote{(2,4)} &
                13.60       \\
            REF-A4           & 
                10722954    &
                red giant branch\tnote{(5)} &
                12.65       \\
            \midrule
            -           &
                9164561    &
                A5V (host star)\tnote{(6)}         &
                14.06       \\
            REF-B1           & 
                9099798     &
                rotationally variable G4V\tnote{(3,4)} &
                13.30       \\
            \bottomrule
        \end{tabular}

        {\bf Notes.} (1) UCAC5 Catalogue \citep{UCAC5}.
        (2) \cite{Bai2018}. (3) \cite{Frasca2016}. (4) \cite{McQuillan2014}. 
        (5) \cite{Stello2013}. (6) \cite{Qian2018}.

        \end{threeparttable}
    \end{table}    
    
    \begin{table}[tbp]
        \renewcommand\arraystretch{1.2}
        \centering
        \begin{threeparttable}[b]
        \caption[]{Observation summary. }
        \label{table_observation}
        \begin{tabular}{lcc}
            \toprule
                         & 
            KIC 10657664 & 
            KIC 9164561 \\
            \midrule
            RA (J2000)          & 
                $\mathrm{19^h 13^m 59.87^s}$   & 
                $\mathrm{19^h 42^m 27.64^s}$    \\
            DEC (J2000)         & 
                $47^{\circ} 59' 47.29''$        & 
                $45^{\circ} 30' 17.00''$        \\
            Starting Date       & 
                16 July 2015    & 
                10 July 2015    \\
            Starting UT         & 
                23:00           & 
                00:10           \\
            Ending UT           & 
                03:30           & 
                05:13           \\
            Exposure time (s)   & 
                30              & 
                100             \\
            Frame number        & 
                334             & 
                153             \\
            Duty cycle (\%)     &  
                58.8            & 
                82.6            \\ 
            Air mass\tnote{(1)}  & 
                1.13 -- 1.06 -- 1.25    & 
                1.09 -- 1.04 -- 1.39    \\
            Seeing\tnote{(2)}& 
                $0.61''$ -- $0.93''$    & 
                $1.49''$ -- $2.28''$    \\
            Resolution \text{\AA}/pix & 
                6.3 -- 9.5 & 
                15.2 -- 23.3 \\
            \bottomrule
        \end{tabular}
        
        {\bf Notes.} (1) Start -- minimum -- end. (2) FWHM of the target PSF along the spatial direction at the central wavelength of 7187 \text{\AA}.

        \end{threeparttable}
    \end{table}

    The spectral data were reduced using the standard IRAF routines together with our customized IDL scripts based on AstroLib\footnote{https://idlastro.gsfc.nasa.gov}. We first performed corrections on overscan, bias, and flat field for all spectral images. We then constructed a two-dimensional pixel-to-wavelength mapping based on the row-by-row line identification of the arc lines. The science spectral images were transformed to the wavelength space so that the sky emission lines are no longer curved in the spatial direction. We masked out all the stars and fit the remaining sky backgrounds with a linear model to predict the variation in sky emission within the stellar point spread function (PSF). Then we transformed the 2D sky models back to the pixel space before subtracting them from the science spectral images. In addition, we performed a simple sigma-clipping method on the time series of each pixel to flag the points that were hit by cosmic rays, which were then replaced by the median values of the neighboring pixels in the corresponding exposures. Finally, we extracted the stellar spectra of the target and the reference stars with the optimal extraction algorithm \citep{Horne1986} through the aperture diameters ranging from 4 to 44 pixels ($1.02''$ -- $11.18''$) with an increment of 1 pixel. Since the target and the reference stars were not perfectly aligned to the central line of the slit and the stars drifted within 1 pixel during the observations, we first applied wavelength correction for each object and each exposure by calculating the wavelength shifts of the H$\alpha$ and Na I lines from the laboratory air wavelengths. We then aligned the telluric absorption lines of the reference star to those of the target star based on the cross-correlation function of spectral profiles around the oxygen A band (759 -- 770 nm).

\begin{figure}[htbp]
        \centering
        \includegraphics[width=\linewidth]{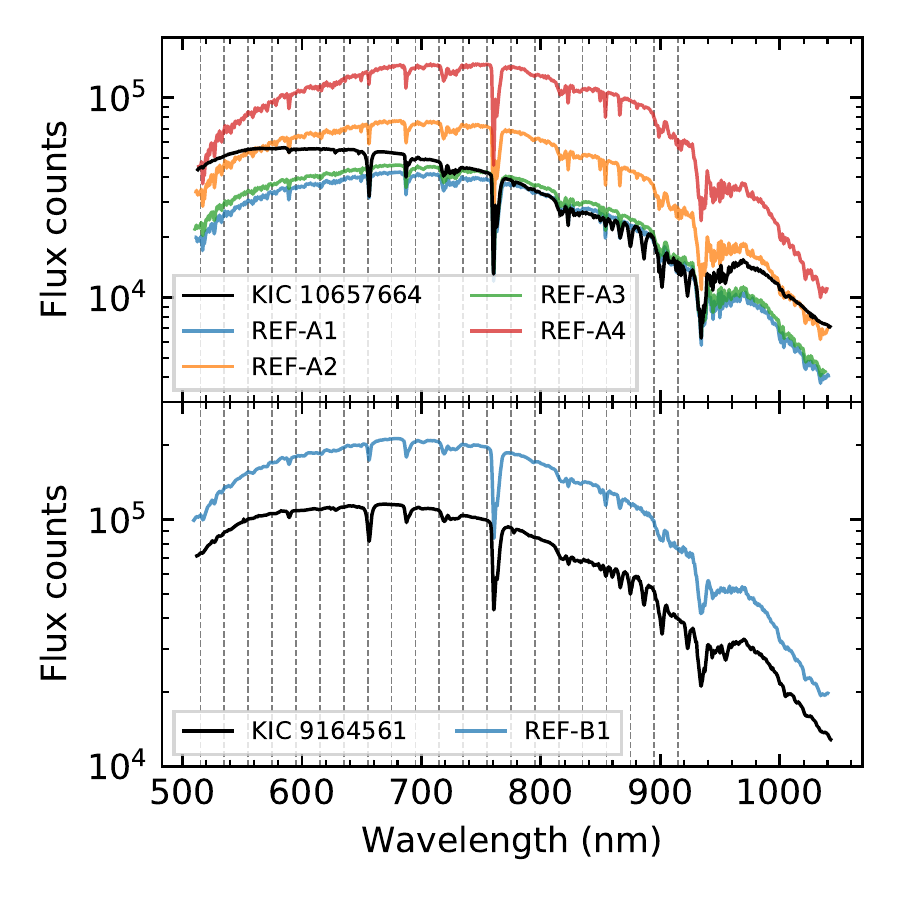}
        \caption{Time-averaged stellar spectra of all objects. 
        The upper panel shows the spectra of KIC 10657664 and its four reference stars. 
        The lower panel shows the spectra of KIC 9164561 and its single reference star.
        The green shaded intervals indicate the wavebands for spectroscopic light curves.}
        \label{fig_stellar_spectra}
    \end{figure}

    Although the R1000R grism was used in combination with a spectral order sorter filter, which cuts out the light blueward from $\sim$495 nm\footnote{ http://www.gtc.iac.es/instruments/osiris/osiris.php\#SecondOrder}, there is a slight second-order contamination in the stellar spectra at 960 -- 980 nm, especially for the two science targets with strong blue radiation. In addition, the fringing in the OSIRIS CCD images are measured to be $\sim$5\% for wavelengths larger than 930 nm. Therefore we only used the wavelength range of 515 -- 915 nm for the light-curve analyses (Fig. \ref{fig_stellar_spectra}). Integrating the spectra in the broadband (excluding 755 -- 775 nm where the telluric oxygen A-band absorption is predominant), we constructed the raw light curves of all the stars. The other telluric absorption bands, including the oxygen B band ($\sim$687 nm) and the water bands (e.g., $\sim$820 nm), were not excluded because they do not have significant impacts on the resulting broadband light curves. The spectroscopic light curves were constructed by integrating the fluxes in uniform 20 nm passbands (Fig. \ref{fig_stellar_spectra}), which are compromised between the signal-to-noise (S/N) ratio and the spectral resolution. We reduced the telluric effects by dividing the light curves of the target star with that of a reference star. The transit light curves were then normalized based on the out-of-transit flux. We determined the best aperture size by minimizing the light-curve scatter, which varies with different pairs of the target and the reference stars. For KIC 10657664, the best aperture diameters are found to be 10 pixels ($2.54''$) when using the REF-A1, REF-A3, and REF-A4 stars, and 26 pixels ($6.60''$) when using the REF-A2 star. For KIC 9164561, the best aperture diameter is 23 pixels ($5.84''$). 

\section{Transit light-curve analysis}
\label{sect_method}

    \subsection{Transit model}
    \label{sect_transit_model}
    
    The modeling of transit light curves is similar to our previous work \citep{2021arXiv210911235J}. We used the Python code \texttt{batman} \citep{code-batman} to calculate the transit models proposed by \cite{Mandel2002}. We adopted the transit ephemeris of \cite{Wong2020} for KIC 10657664 and those of \citet{Rappaport2015} for KIC 9164561 to roughly predict the central transit time assuming circular orbits. We used the quadratic stellar limb-darkening law in the transit model. The limb-darkening coefficients $u_1$ and $u_2$ were constrained by Gaussian priors, where the prior means were calculated with the code developed by \cite{code-limb-darkening} based on the \texttt{ATLAS9} model atmospheres \citep{code-ATLAS9}, and the prior uncertainties were assumed to be 0.1. For broadband light-curve modeling, the free parameters in a transit model include the companion-to-host radius ratio $R_2/R_1$, the quadratic limb-darkening coefficients $u_1$ and $u_2$, the semimajor axis relative to the host star radius $a/R_1$, the orbital inclination $i$, and the central transit time $t_{\rm c}$. The spectroscopic light curves were separately fit in each narrow band. Since the orbital parameters ($a/R_1$, $i$, and $t_{\rm c}$) are independent of wavelength, we fixed these parameters to the median estimates derived from white-light curve fitting. Thus the free parameters used for spectroscopic light-curve fitting are $R_2/R_1$, $u_1$, and $u_2$. 
    
    The nightside thermal emission of most exoplanets is negligible with current instrumental precision, except for some ultrahot Jupiters observed in infrared wavebands \citep[e.g.,][]{Kipping2010, Chakrabarty2020}. In contrast, the self-emission from white dwarfs can significantly dilute the transit signals in the optical waveband, which would cause an underestimate of the transit depth if not corrected. The diluted transit model is calculated to be 
    \begin{equation}\label{eq_dilution}
        m^*(t; \mathcal{F}) = \frac{m(t) + \mathcal{F}} {1 + \mathcal{F}},
    \end{equation}
    where $m(t)$ is the transit model without the dilution effect, and $\mathcal{F}$ is the companion-to-host flux ratio in the corresponding waveband. The flux ratio $\mathcal{F}$ can be directly measured through secondary eclipse observations, but we lack such spectroscopic measurements covering the same wavebands. The flux ratio varies with wavelength and is degenerate with the transit depth, for which we cannot take the estimates from the {\it Kepler} data, nor can we directly retrieve it from the GTC spectroscopic light curves. Therefore we estimated the spectroscopic flux ratios using the blackbody assumption,
    \begin{equation}\label{eq_flux_ratio}
        \mathcal{F} (R_2/R_1, T_1, T_2; \lambda_a, \lambda_b)
        = \left(\frac{R_2}{R_1}\right)^2\cdot
          \frac{\int_{\lambda_{a}}^{\lambda_{\rm b}} B(\lambda, T_2)\mathrm{d}\lambda}
               {\int_{\lambda_{a}}^{\lambda_{\rm b}} B(\lambda, T_1)\mathrm{d}\lambda},
    \end{equation}
    where $R_2/R_1$ is the radius ratio, $T$ is the effective temperature, the subscripts 1 and 2 refer to the primary and the secondary star, respectively, $(\lambda_a, \lambda_b)$ is the wavelength interval, and $B(\lambda, T)$ is the Planck function. We adopted the temperature estimates from \cite{Wong2020} and \cite{Rappaport2015}, which are $T_1=9940^{+260}_{-230}$ K and $T_2=15\,080\pm400$ K for the KIC 10657664 system, and $T_1=7870\pm150$ K and $T_2=10\,410\pm200$ K for the KIC 9164561 system. We find that the measurement uncertainties of $T_1$ and $T_2$ have negligible impacts on the light-curve fitting. Therefore we used fixed values instead of assuming Gaussian priors on $T_1$ and $T_2$ so as to reduce the number of free parameters in the transit model.

\subsection{Baseline variation effects in binary systems}

    The effects of Doppler boosting (DB), mutual illumination (ILL), and ellipsoidal light variation (ELV) may introduce additional periodic features in the phase curve of a binary system, and thus might change the baseline of transit light curves. Both \cite{Wong2020} and \cite{Rappaport2015} used third-order harmonic series to account for these effects when analyzing the {\it Kepler} light curves. As introduced in \cite{Carter2011}, the features of DB, ILL, and ELV mainly correspond to the $\sin(\phi)$, $\cos(\phi)$, and $\cos(2\phi)$ terms, respectively, where $\phi(t)=2\pi(t-t_{\rm c})/P$ is the orbital phase and $P$ is the orbital period. Based on the results of \cite{Wong2020} and \cite{Rappaport2015}, we can roughly evaluate the baseline variation amplitudes (BVA) induced by these systematic effects during a five-hour transit observation with the GTC OSIRIS instrument. The total astrophysical induced BVA is $\sim$40 ppm for KIC 10657664, which is negligible compared to the photometric precision of $\sim$200 ppm. Therefore we did not consider these astrophysical effects when fitting the transit light curves of KIC 10657664. For the other system, KIC 9164561, although the BVA due to the DB effect is only $\sim$10 ppm, those due to the ILL and ELV effects can reach considerable amplitudes of $\sim$300 ppm and $\sim$4000 ppm in the GTC transit observation. This is because the amplitudes of ILL obeys the inverse square law and that of ELV is inversely proportional to the fourth power of the relative orbital distance. Since the value of $a/R_1$ for KIC 9164561 ($a/R_1\approx2.5$) is considerably lower than that for KIC 10657664 ($a/R_1\approx7.0$), the effects of ILL and ELV are consequently much stronger for the latter. 
    Thus the significant baseline variations should not be ignored for KIC 9164561. 
    
    We corrected for the effects of ILL and ELV in the light-curve analyses of KIC 9164561 by multiplying its transit model with a harmonic series, 
    \begin{equation} \label{eq_harmonic_correction}
        h(\phi)=1+A_{\rm ILL}\cos(\phi)+A_{\rm ELV}\cos(2\phi),
    \end{equation}
    where the harmonic amplitudes $A_{\rm ILL}$ and $A_{\rm ELV}$ are free parameters. The other harmonic terms with small amplitudes, such as $A_{\rm DB}\sin(\phi)$, $A_{\rm ELV}\cos(\phi)$, and $A_{\rm ILL}\cos(2\phi)$, are not included in Eq. \ref{eq_harmonic_correction} because they may be treated as noise and be accounted by the noise models illustrated in Sect. \ref{sect_noise_model}. We took $A_{\rm ILL} \sim \mathcal{N}(2385, 36)$ (in unit of ppm) from \cite{Rappaport2015} for all the broadband and spectroscopic light curves of KIC 9164561. We do not expect a significant wavelength dependence of $A_{\rm ILL}$ according to the ILL model presented in \cite{Carter2011}. We note that the ELV model depends on the host star limb-darkening and gravity-darkening effects \citep{Morris1985, Carter2011}. Thus the wavelength dependency of $A_{\rm ELV}$ should be considered when fitting spectroscopic light curves. According to Eqs. (2) and (3) in \cite{Morris1985}, the amplitude of ELV is predicted to be 
    \begin{equation} \label{eq_elv}
        A_{\rm ELV} \approx -\frac{45+3u}{20(3-u)}(1+\tau) q (R_1/a)^3 \sin^2{i},
    \end{equation}
    where $u$ is the linear limb-darkening coefficient, $\tau$ is the gravity-darkening coefficient, $q$ is the companion-to-host mass ratio, $R_1/a$ is the host star radius relative to the orbital semimajor axis, and $i$ is the orbital inclination. Taking $q=0.097\pm0.004$, $R_1/a=0.397\pm0.003$, and $i=71.59\pm0.22~\rm{deg}$ from \cite{Rappaport2015}, we need to calculate $u$ and $\tau$ in each waveband so as to obtain the prior constraints on $A_{\rm ELV}$. The linear limb-darkening coefficients were calculated with the code of \cite{code-limb-darkening} based on the stellar ATLAS model \citep{Kurucz1979}, which we previously used to calculate the quadratic limb-darkening coefficients for the transit model. The gravity-darkening coefficients were calculated with Eq. (10) in \cite{Morris1985}. The predicted values of $A_{\rm ELV}$ in each narrowband are listed in Table \ref{table_prior_elv}, and their estimated uncertainties are $\sim$400 ppm according to error propagation of Eq. \ref{eq_elv}. We calculated the weighted means of $u$ and $\tau$ with the instrumental response curve of the OSIRIS R1000R grism as the weights to obtain $u=0.43$, $\tau=0.72$, and $A_{\rm ELV}=-8466\pm400$ ppm in the broadband, which is slightly lower than the measured value of $A_{\rm ELV}=-8880\pm35$ ppm from \cite{Rappaport2015} because the {\it Kepler} waveband covers shorter visible wavelengths starting from 420 nm.

    \subsection{Noise model}
    \label{sect_noise_model}

    There are typically two ways to handle light-curve systematics. One way is to use nonparametric stochastic models such as Gaussian process (GP) regression introduced in \cite{Gibson2012} and \cite{code-celerite}. A GP model estimates the correlation between the input variables and the systematics by constructing a covariance matrix with a kernel function, which provides an adaptive and robust characterization of systematic noise. The other way is to use parametric deterministic models, where the systematics are accounted for by a set of baseline functions (BFs). There are various forms of baselines that can be defined, for example, a polynomial series of the state vectors or a harmonic series of time. Exponential baselines have been used to analyze the light curves of the {\it Hubble} Space Telescope \citep[e.g., ][]{Berta2012ApJ, Kreidberg2014Nature}. The ``best'' baseline model needs to be selected from a large number of presumed models and depends on the specific observational conditions and instruments. \cite{Gibson2014} tested these two methods using simulated transit light curves. He suggested that multiple techniques should always be used to test whether the inference is dependent on subjective choices made. Here we focus on the performance of these methods on real data from ground-based observations and investigate the model dependence of the derived transmission spectra.

    \subsubsection{Using Gaussian processes} 
    \label{sect_gp} 
    
    We use GP regression to account for the correlated noise in transit light curves, which was implemented with the \texttt{celerite} code developed by \cite{code-celerite}. The kernels supported by \texttt{celerite} are mixtures of exponentials, which enable fast modeling of one-dimensional GPs. A detailed introduction to the GP framework is available in \cite{Rasmussen2006}. \cite{Gibson2012} and \cite{Gibson2013a,Gibson2013b} introduced and discussed the specific applications of GP modeling in transit light-curve analyses. 

    In a GP model, the observed light curve $\bm f$ follows a multivariate Gaussian distribution,
    \begin{equation}
            \bm f \sim \mathcal{N} 
                ({\bm m}^*(\bm t; \bm\theta), 
                \bm K (\bm t; \bm\varphi)),
    \end{equation}
    where $\bm m^*$ is the mean function with a parameter vector $\bm\theta$, $\bm K$ is the covariance matrix with a parameter vector $\bm\varphi$, and $\bm t$ is the time vector. The mean function $\bm m^*$ is exactly the diluted transit model defined in Eq. \ref{eq_dilution}. The covariance matrix $\bm K$ characterizes the residual noise $\bm r = \bm f - \bm m^*$ consisting of correlated noise and white noise. The elements in $\bm K$ are determined by a covariance function (also called kernel function) $k(\tau_{ij}; \bm{\varphi})$, where $\tau_{ij}\equiv |t_i-t_j|$ is the time difference between two data points.

    In this work we test two well-defined kernels. One is the 3/2-order Mat\'ern kernel (hereafter the M32 kernel), which takes the form
    \begin{equation} \label{eq_Matern32Term}
    \begin{split}
        &k(\tau_{ij}; \bm\varphi) = 
            \alpha^2 \left(1 + \frac{\sqrt{3}\tau_{ij}}{\ell}\right) 
            \exp{\left(- \frac{\sqrt{3}\tau_{ij}}{\ell}\right)}, \\
        &\bm\varphi = (\alpha, \ell),
    \end{split}
    \end{equation}     
    where $\alpha$ and $\ell$ are the amplitude and length scale of the correlated noise. This kernel is approximated with the Mat\'ern32 term in \texttt{celerite}. The other kernel function we test is a stochastically driven damped simple harmonic oscillator (hereafter the SHO kernel). It is computationally efficient for modeling stellar variations \citep{code-celerite}. The SHO kernel exhibits different properties depending on its quality factor $Q$. In the small $Q$ limit ($0 < Q \ll 1/2$), it is over damping and presents no oscillatory features, while in the large $Q$ limit ($Q \gg 1$), it becomes 
    \begin{equation} \label{eq_SHOTerm}
    \begin{split}
        &k(\tau_{ij}; \bm\varphi) = 
            S_0 Q \omega_0 
            \exp{\left(- \frac{\omega_0 \tau_{ij}}{2Q}\right)} 
            \cos{(\omega_0 \tau_{ij})}, \\
        &\bm\varphi = (S_0, Q, \omega_0),
    \end{split}
    \end{equation}     
    where $S_0$ controls the amplitude of the oscillator, and $\omega_0$ is the characteristic frequency. This kernel can be used to characterize the asteroseismic oscillations \citep[e.g.,][]{Grunblatt2017, Pereira2019}.
    
    To account for the white noise in light curves, we added a diagonal jitters term to the kernel function such that
    \begin{equation} \label{eq_add_jitters}
    \begin{split}
        &k'(\tau_{ij}; \bm\varphi') = 
            \varepsilon^2 \delta_{ij} + k(\tau_{ij}; \bm\varphi), \\
        &\varepsilon^2 = \varepsilon_{ii}^2+\varepsilon_{\rm ex}^2, \\
        &\bm\varphi' = (\varepsilon_{\rm ex}, \varphi_1, ..., \varphi_n),
    \end{split}
    \end{equation}      
    where $\delta_{ij}$ is the Kronecker delta, $\varepsilon$ is the total white noise, $\varepsilon_{ii}$ is the flux uncertainty propagated from photon-dominated noise, and $\varepsilon_{\rm ex}$ is a factor to account for white-noise underestimation.

    \subsubsection{Using parametric baseline functions}
    
    When using parametric baseline functions $b(t; \bm\varphi)$, the transit light curves are fit by 
    \begin{equation}
        f(t) = m^*(t; \bm \theta) b(t; \bm\varphi) + \varepsilon,
    \end{equation}
    where $\bm\varphi$ is the parameter vector of the baseline, and $\varepsilon$ is the total white noise (Eqs. \ref{eq_add_jitters}). Different from the GP models, usually many similar BF models can be enumerated, and model selection must be conducted to determine the best model. In this context, we compared the Bayesian evidence among all the BF models when fitting white-light curves and selected the one with the highest evidence as the best model that we used in the subsequent spectroscopic light curve analyses.
    
    For KIC 10657664, the systematics are mainly instrumental and telluric origins. We define the polynomials of state vectors up to the third order as the BF models. We used six state vectors, which are the time $\bm t$, the target position drift $\bm x$ along the dispersion direction and $\bm y$ along the spatial direction, the full width at half maximum (FWHM) of the target PSF in the spatial direction $\bm w$, the instrumental rotation angle $\bm a$, and the airmass $\bm z$, to build a total of 4096 ($=4^6$) BF models. For instance, one baseline function can be written as $b(t) = c_0 + \sum_{n=1}^{3}{c_{t,n} t^n} + \sum_{n=1}^{2}{c_{w,n} w^n(t)}$, and $\bm\varphi=(c_0, c_{t,1}, c_{t,2}, c_{t,3}, c_{w,1}, c_{w,2})$. The state vectors were normalized by their maximum norms before being added into polynomials. Fig. \ref{fig_state_vectors} shows the time series of the original state vectors.
    
    For KIC 9164561, the systematics are mainly due to stellar pulsations. We find that the polynomial baselines composed of state vectors are unsuitable for its light-curve fitting. Therefore we followed \cite{Essen2019} and used the sum of sinusoidal functions to account for the oscillatory systematics. The sinusoidal BF for the light curves of KIC 9164561 has the form
    \begin{equation} \label{eq_sinusoidal_bf}
        b(t) = c_0 + \sum_{i=1}^{m} A_i \sin(2\pi t/T_i - \phi_i),
    \end{equation}
    where $A_i$ is the amplitude, $T_i$ is the period, $\phi_i$ is the phase, and $m$ is the number of terms. The prior constraints on each sinusoidal term were assumed based on the periodogram of the white-light curve, while $c_0$ is assumed to follow $\mathcal{U}(0.5, 1.5)$.

    \subsubsection{Reducing common-mode systematics}
    
    A part of the systematics in spectroscopic light curves is instrument induced and does not vary with wavelength. This is the so-called ``common-mode'' systematics. Reducing the common modes when fitting spectroscopic light curves can lower the noise level and thus improve the S/N. One example is the \texttt{divide-white} method outlined in \cite{Gibson2013b} and \cite{Stevenson2014}, where the common-mode systematics are approximated with the white-light systematics $W(t) = f(t) / m^*(t)$. The spectroscopic light curves are then divided by the white-light systematics,
    \begin{equation} \label{eq_divide_white}
        f^*_\lambda(t) = \frac{f_\lambda(t)}{W(t)} ,
    \end{equation}
    where $f_\lambda(t)$ is the normalized spectroscopic light curve observed at wavelength $\lambda$. Since the broadband systematics can be be viewed as the average of narrowband systematics weighted by flux, $W(t)$ is a good approximation to the common-mode systematics if the systematics in different narrowbands have weak wavelength dependences. 
    
    However, the spectroscopic systematics may be largely dependent on wavelength (or flux) if they are mainly induced by the astrophysical processes such as stellar activities or pulsations of either the target or the reference star. In this scenario, the \texttt{divide-white} method might incorrectly estimate the noise level in different wavebands, causing unexpected modification in some narrowband light curves. Therefore, in addition to the \texttt{divide-white} method, we also tested whether it would present a better performance if a scaling factor $\eta_\lambda$ were added as a free parameter to account for the wavelength dependence of spectroscopic systematics. A similar method has been outlined in \cite{Mandell2013}. We call this variant method as the \texttt{divide-rescaled} method, where the spectroscopic light curve is corrected to be
    \begin{equation} \label{eq_divide_variable} 
        f^*_\lambda(t) = \frac{f_\lambda(t)}{1 + \eta_\lambda (W(t)-1)}.
    \end{equation}
    When $\eta_\lambda=1$, the above equation is equivalent to Eq. \ref{eq_divide_white}, while $\eta_\lambda$ approaches zero if the light curve at wavelength $\lambda$ does not share the common-mode patterns.
    
    In addition to these two methods, we also tried to fit the spectroscopic light curves directly without reducing the common-mode systematics, which is denoted as the \texttt{direct-fit} method. It serves as a null model such that we can investigate how a large amplitude of common-mode systematics would impact the transmission spectrum.

    \subsection{Bayesian estimation and model comparison}
    \label{sect_bayesian_framework}
    
    The commonly used methods for Bayesian parameter estimation include the Markov chain Monte Carlo method (e.g., the Metropolis-Hastings algorithm, \citealt{Metropolis1953,Hastings1970}; and the Hamiltonian Monte Carlo algorithm, \citealt{Betancourt2017}) and nested sampling \citep{Skilling2004,Feroz2008}. The MCMC method is highly effective for obtaining a stable estimate of the posterior joint distributions of the parameters. The nested sampling algorithm can also obtain the posterior estimates of the parameters, although it is initially designed for computing Bayesian evidence (also called the marginalized likelihood), which provides a criterion for model comparison. Since model comparison is a critical part in this work, we adopted the nested sampling method to obtain the Bayesian inferences. This method uses a set of sampling points to fully explore the prior space from the prior boundary toward higher-likelihood regions and is able to reveal multimodal distributions when implemented with the \texttt{MULTINEST} algorithm \citep{code-MultiNest}. The sampling naturally comes to a convergence when the remaining parameter space has negligible evidence contributions. When the sampling is complete, we obtain both Bayesian evidence and the posterior estimates.

    We used \texttt{PyMultiNest} \citep{code-PyMultiNest}, which is built on the \texttt{MULTINEST} library, to estimate Bayesian evidence and parameter posteriors. In the context of Bayes' rule, the model evidence $\mathcal{Z}$ can be calculated by integrating the likelihood over the prior space of parameters $\bm \Theta$, which intrinsically is the average likelihood of a model hypothesis over its prior space. Following \cite{Feroz2008}, this can be written as
    \begin{equation}
        \mathcal{Z}(\mathcal{D} | \mathcal{H}) = \int \mathcal{L}(\mathcal{D}|\bm \Theta, \mathcal{H})
            \pi(\bm \Theta|\mathcal{H}) \mathrm{d}^D \bm \Theta,
    \end{equation}
    where $\mathcal{D}$ is the data, $\mathcal{H}$ is the model hypothesis, $\mathcal{L}$ is the likelihood function, $\pi$ is the prior function, and $D$ is the dimensionality of the parameter space. When the data are given, the model hypothesis with a lower dimensionality but a higher maximum likelihood usually presents stronger Bayesian evidence and thus prevails in model comparison.
    
    In practice, we acquired $\ln\mathcal{Z}$ by calculating the natural logarithmic likelihood,
    \begin{equation}
        \ln\mathcal{L} = -\frac{N}{2}\ln{(2\pi)} 
            - \frac{1}{2} \ln{|\bm K|} 
            - \frac{1}{2} \bm r^T \bm K^{-1} \bm r,
    \end{equation}    
    where $N$ is the number of data points. The residual vector $\bm r$ and the covariance matrix $\bm K$ have been defined in Sect. \ref{sect_gp} for the GP models. When using the BF models, the residual is $r(t) = f(t) - m^*(t)b(t)$, and $\bm K$ is just a diagonal matrix composed of jitter terms. We estimated $\ln\mathcal{Z}$ with 1000 sampling points, a sampling efficiency of 0.3, and an evidence tolerance of 0.1 when running \texttt{PyMultiNest}. The corresponding precision of $\ln\mathcal{Z}$ is 0.1 -- 0.5, depending on model complexity. 
    
    We compared the model by comparing $\ln\mathcal{Z}$ between two models,  which is equivalent to the natural logarithmic Bayes factor, 
    \begin{equation}
    \ln \frac{\mathrm{pr}(\mathcal{D} | \mathcal{H}_1)}{\mathrm{pr}(\mathcal{D} | \mathcal{H}_2)} 
    = \ln \left(\frac{\mathcal{Z}_1}{\mathcal{Z}_2}\right),
    \end{equation}
    where the subscripts 1 and 2 denote hypothesis 1 ($\mathcal{H}_1$) and hypothesis 2 ($\mathcal{H}_2$), respectively. We adopted the categories proposed by \cite{Kass1995} that $\mathcal{H}_1$ has very strong evidence against $\mathcal{H}_2$ if $\ln(\mathcal{Z}_1 / \mathcal{Z}_2) \ge 5$; $\mathcal{H}_1$ has strong evidence against $\mathcal{H}_2$ if $3 \le \ln(\mathcal{Z}_1 / \mathcal{Z}_2) < 5$; $\mathcal{H}_1$ has positive evidence against $\mathcal{H}_2$ if $1 \le \ln(\mathcal{Z}_1 / \mathcal{Z}_2) < 3$; and two hypotheses have comparable evidence if $|\ln(\mathcal{Z}_1 / \mathcal{Z}_2)| < 1$.

\section{Results and discussion}
\label{sect_results}

    \subsection{White-light curve fitting}
    
    \begin{figure*}[htbp]
        \centering
        \includegraphics[width=\linewidth]{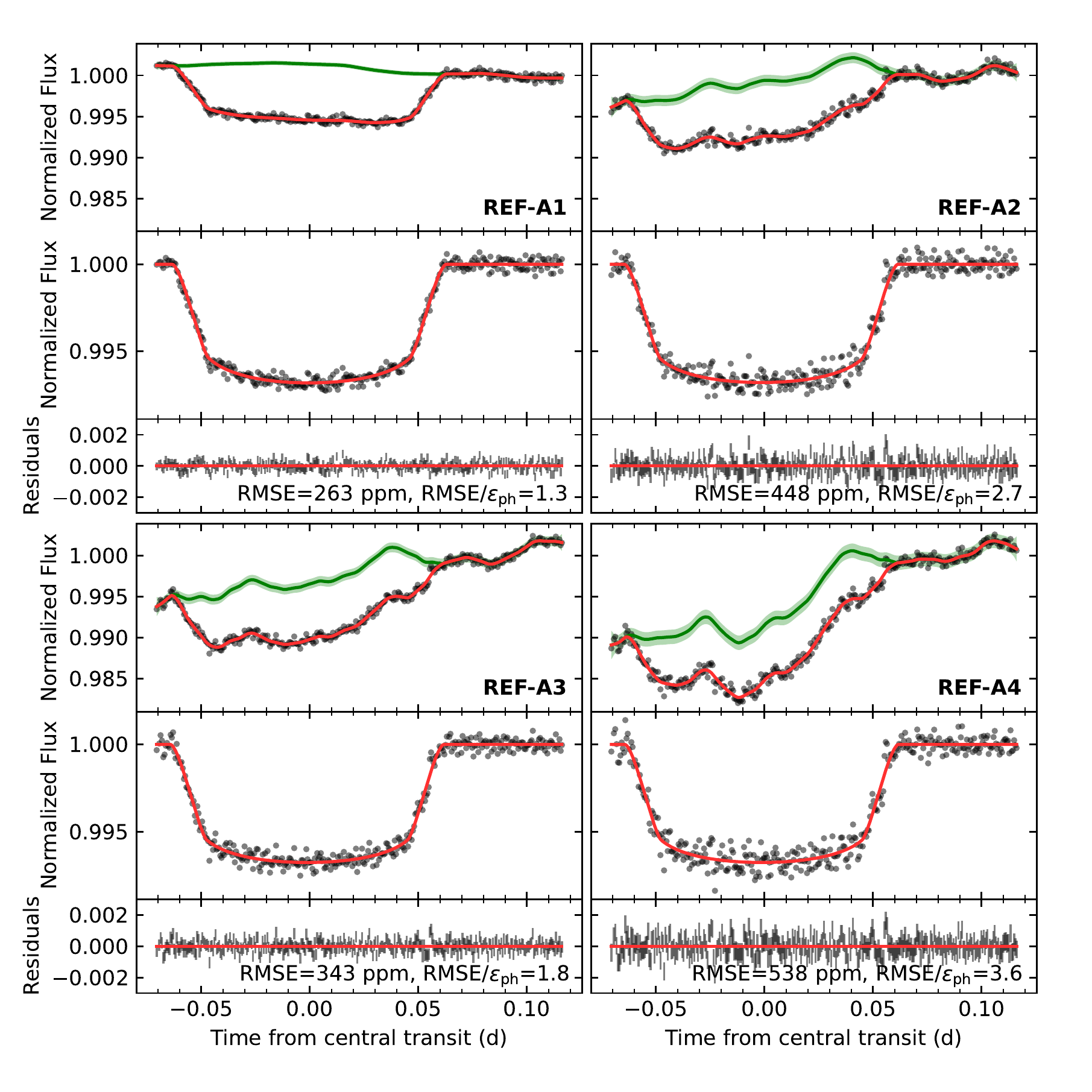}
        \caption{Results of white-light curve fitting for KIC 10657664 using the GP-M32 model. The four split parts correspond to the light curves calibrated with four reference stars (REF-A1 to A4). In each part, the top panel shows the white-light curve (black dots), the best-fit model (red), and the GP systematics (green), the middle panel shows the detrended light curve (black dots) and the transit model (red), and the bottom panel shows the residuals, where $\varepsilon_{\rm ph}$ denotes the uncertainty of photon-dominated noise. The green shaded intervals indicate 5$\sigma$ confidence of the GP models. The GP systematics are shifted to match the light-curve baselines.}
        \label{fig_white_10657664}
    \end{figure*}
    
    \begin{figure*}[htbp]
    \centering
    \includegraphics[width=\linewidth]{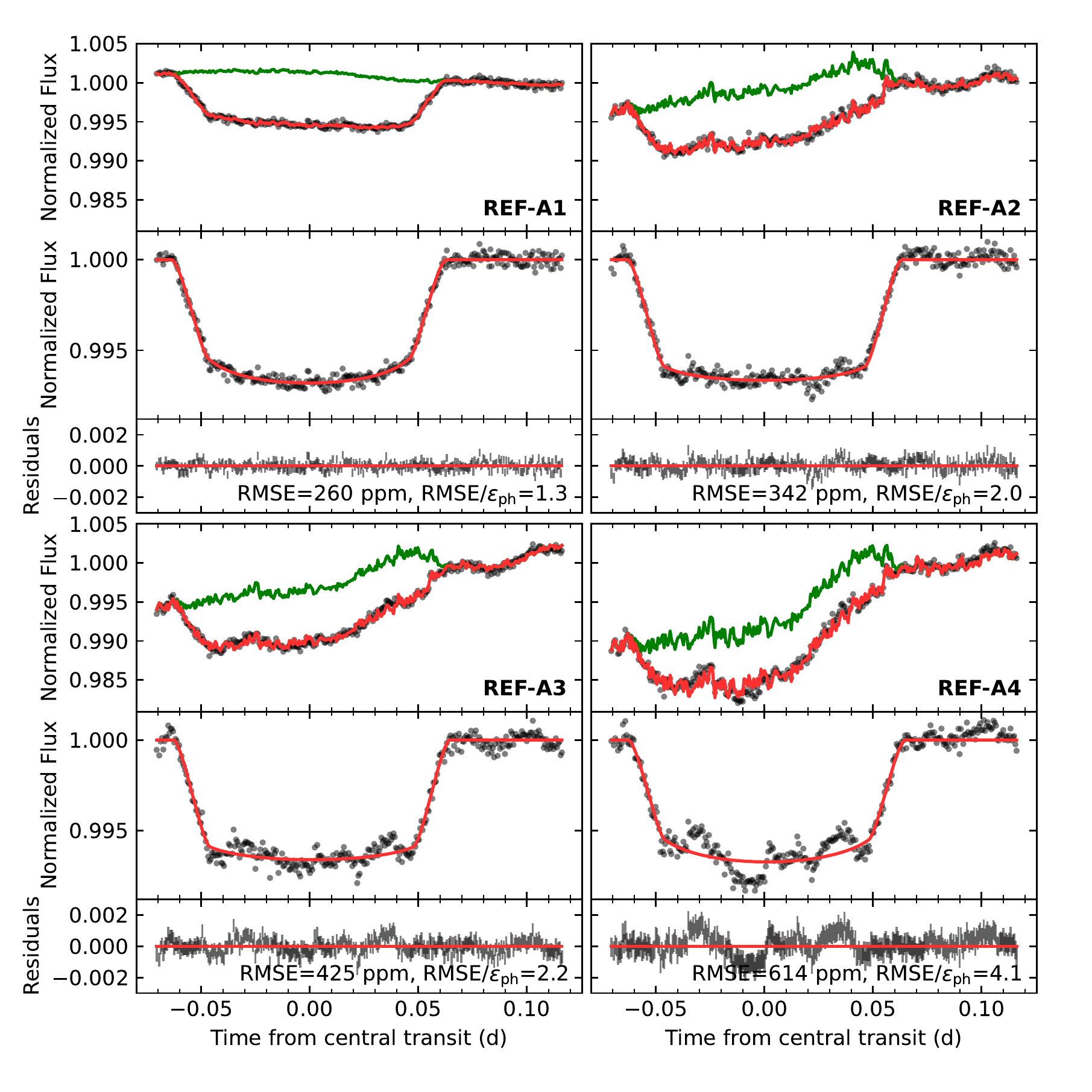}
    \caption{Same as Fig. \ref{fig_white_10657664}, but the systematics are modeled with the best-fit BFs (\#1) listed in Table \ref{table_bf_evidence_1065}.}
    \label{fig_white_10657664_BF}
    \end{figure*}
    
    \begin{figure*}[htbp]
        \centering
        \includegraphics[width=\linewidth]{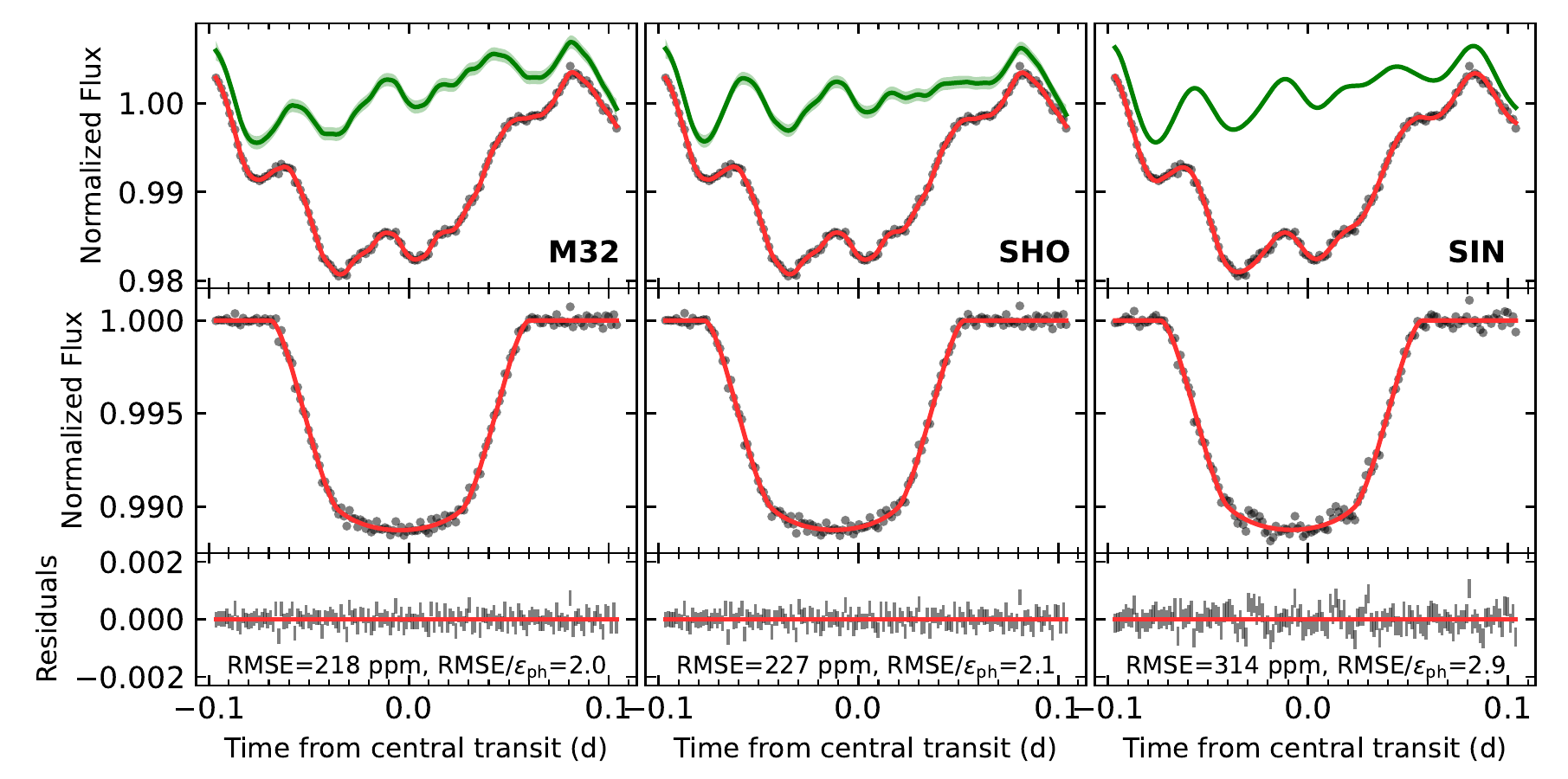}
        \caption{Results of white-light curve fitting for KIC 9164561 using different noise models.
        {\it Left column}: GP-M32 model. {\it Middle column}: GP-SHO model. {\it Right column}: Sinusoidal BF model. {\it Top row}: White-light curve (black dots), best-fit model (red), and systematic noise (green). {\it Middle row}: Detrended light curves (black dots) and transit model (red). {\it Bottom row}: Residuals. The green shaded intervals indicate the 5$\sigma$ confidence of the GP models. The GP systematics are shifted to match the light-curve baselines. The systematics exhibit positive offsets due to the ILL and ELV effects.}
        \label{fig_white_9164561}
    \end{figure*}

    Figures \ref{fig_white_10657664} and \ref{fig_white_10657664_BF} show the best-fit white-light curves of KIC 10657664 calibrated with four different reference stars, where the systematics are modeled with GPs and BFs, respectively. It is evident that the amplitudes of light-curve systematics vary with the adopted reference star. The light curve calibrated with REF-A1 exhibits the weakest systematics among the four groups of results, and its root-mean-squared error (RMSE) reaches $\sim$1.3 times photon noise when fit with GP models. The other three light curves show stronger systematics with similar patterns. We speculate that the difference of systematics in amplitudes is mainly related to the distance between the star position and the rotation center of the OSIRIS field of view. As shown in Fig. \ref{fig_slit}, the three reference stars (REF-A2, REF-A3, and REF-A4) placed on CCD2 are farther away from the rotation center compared with the target star and the REF-A1 star. A similar scenario was reported in \cite{Nortmann2016}, where the bump features in systematic noise might be attributed to vignetting in pupil space due to instrument rotation.  Additionally, it is impossible to keep the five objects perfectly aligned with the central line of the long slit ($7' \times 12''$) during the observation. Therefore, discrepant flux variations might result from varying pixels or slight differential light losses produced by the finite slit width. We also tested whether the larger-amplitude systematics for REF-A2 to REF-A4 could be induced by differential sky variability, but found no correlation between them. In addition, the variability of the reference stars cannot explain the similarity of systematics in the light curves calibrated with REF-A2, REF-A3, and REF-A4 either.
    
    The estimated transit parameters of KIC 10657664B are listed in Table \ref{table_transit_parameters_1065}. We tested light-curve fitting assuming uninformative priors, but found that the transit parameters, particularly $R_2/R_1$, $a/R_1$, and $i$, failed to be constrained by the transit light curves using REF-A2, REF-A3, and REF-A4. Therefore we adopted Gaussian priors based on the estimates of \cite{Wong2020}, which are well constrained by the collection of all 18 quarters of the {\it Kepler} long-cadence light curves. We used the transit depth obtained in the {\it Kepler} band (420 -- 900 nm) to constrain that obtained in the OSIRIS {\it R} band (515 -- 915 nm) assuming that the transit depths of white dwarfs are independent of wavelength. Furthermore, our white-light curve analyses serve to extract precise and reliable broadband systematic noise, instead of improving the transit parameter estimation, which is critical for removing the common-mode systematics in the spectroscopic light curves.
    
    The case of KIC 9164561 is more complicated. We used Gaussian priors to constrain $R_2/R_1$, $a/R_1$, and $i$ based on the estimates of \cite{Rappaport2015} (18 {\it Kepler} quarters). Otherwise, the transit signals could not be correctly distinguished from the correlated noise through GP modeling because of the strong stellar pulsations and the short observation time out of transit. As described in Sect. \ref{sect_transit_model}, we additionally considered the effects of ELV and ILL when fitting the light curves of KIC 9164561. The best-fit white-light curves are shown in Fig. \ref{fig_white_9164561}, and the corresponding transit parameters are listed in Table \ref{table_transit_parameters_9164}. According to the results, we find that the M32 and the SHO kernels predict different systematics and central transit time ($t_\mathrm{c}=-0.00331^{+0.00438}_{-0.00512}~\mathrm{d}$ when using the M32 kernel; $t_\mathrm{c}=-0.01426^{+0.00171}_{-0.00157}~\mathrm{d}$ when using the SHO kernel; Table \ref{table_transit_parameters_9164}). These discrepancies mainly arise because the boundaries of transit ingress and egress become indistinct due to the large amplitudes of oscillatory noise. We did not assume a Gaussian prior on $t_{\rm c}$ because the companion white dwarf exhibits a long-term transit timing variation reported by \cite{Rappaport2015}, which might arise from an undiscovered third body with an orbital period of 8 to 14 years. Because the GTC transit observation is $\sim$790 days ($\sim$623 binary orbits) later than the last transit observed by {\it Kepler}, it is difficult to tell which is closer to the true value, the M32-derived $t_{\rm c}$ or the SHO-derived $t_{\rm c}$. Therefore, we kept both results derived from the two GP models and separately used their systematics for common-mode corrections in the subsequent spectroscopic light-curve analyses of KIC 9164561.

    In addition to using GP models, we also tried to use various BF models to account for light-curve systematics. We conducted a Bayesian model comparison among 4096 different BFs for KIC 10657664 and eight sinusoidal models for KIC 9164561 based on their white-light curves. In the subsequent spectroscopic light curve fitting, we adopted those BFs with the strongest Bayesian evidence. The results of Bayesian evidence estimated by the nested sampling algorithm are listed in Table~\ref{table_bf_evidence_1065} for KIC 10657664 and Table~\ref{table_bf_evidence_9164} for KIC 9164561, where the evidence for GP models is also listed for comparison. Table~\ref{table_bf_evidence_1065} shows that the best BF model changes with the adopted reference star, while all of the best models show correlations with the FWHM of the stellar PSF and the rotation angle of the instrument, suggesting the sources of light curve systematics. However, as shown in Fig.~\ref{fig_white_10657664_BF}, the best BFs selected from a large number of models still have a poor performance in characterizing the light-curve systematics of REF-A3 and REF-A4. On the other hand, Table~\ref{table_bf_evidence_9164} shows the best BF for KIC 9164561 consists of eight sinusoidal terms, but the corresponding fitting result is still worse than those of the GP models (Fig.~\ref{fig_white_9164561}). We note that the Bayesian evidence of the sinusoidal BF continues to increase with the number of harmonic terms, but we did not use more complex models with nine or more harmonic terms because their computational cost is high. For instance, when nine harmonic terms are used in Eq.~\ref{eq_sinusoidal_bf}, it $\text{takes about}$10 hours for a 64-core parallel computation (3 GHz CPUs) to fit one light curve. Meanwhile, adding more trivial terms with higher frequencies will not significantly improve the results. Therefore, we adopted $m=8$ in Eq.~\ref{eq_sinusoidal_bf} in the subsequent light-curve analyses of KIC 9164561.

    \subsection{Comparison of transmission spectra}
    \label{sect_spectra_comparison}

    The raw spectroscopic light curves of KIC 10657664 and KIC 9164561 are shown in Figure \ref{fig_spec_lightcurves}. We used the methods described in Sect. \ref{sect_method} to compute the transmission spectra of the two targets. We assumed uniform priors of $\mathcal{U}(0.03, 0.3)$ on narrowband transit depths when fitting the spectroscopic light curves. We did not assume Gaussian priors on $R_2/R_1$ in spectroscopic light-curve fitting so as to avoid the resulting transmission spectra being largely affected by the informative priors. We compared the computed transmission spectra with the broadband transit depths to check the general offsets of transit depth. We analyzed the potential wavelength dependency of a transmission spectrum by fitting the spectrum with polynomials from the zeroth order up to the 19th order and then selected the best-fit polynomial with the strongest model evidence. A transmission spectrum is determined to be flat and featureless if the zeroth-order polynomial (i.e., a constant function) presents the strongest evidence. Otherwise, the transmission spectrum is determined to exhibit spurious spectral features resulting from data analysis procedures or noise fluctuation.

    \subsubsection{KIC 10657664B}

\begin{figure*}[htbp]
        \centering
        \includegraphics[width=\linewidth]{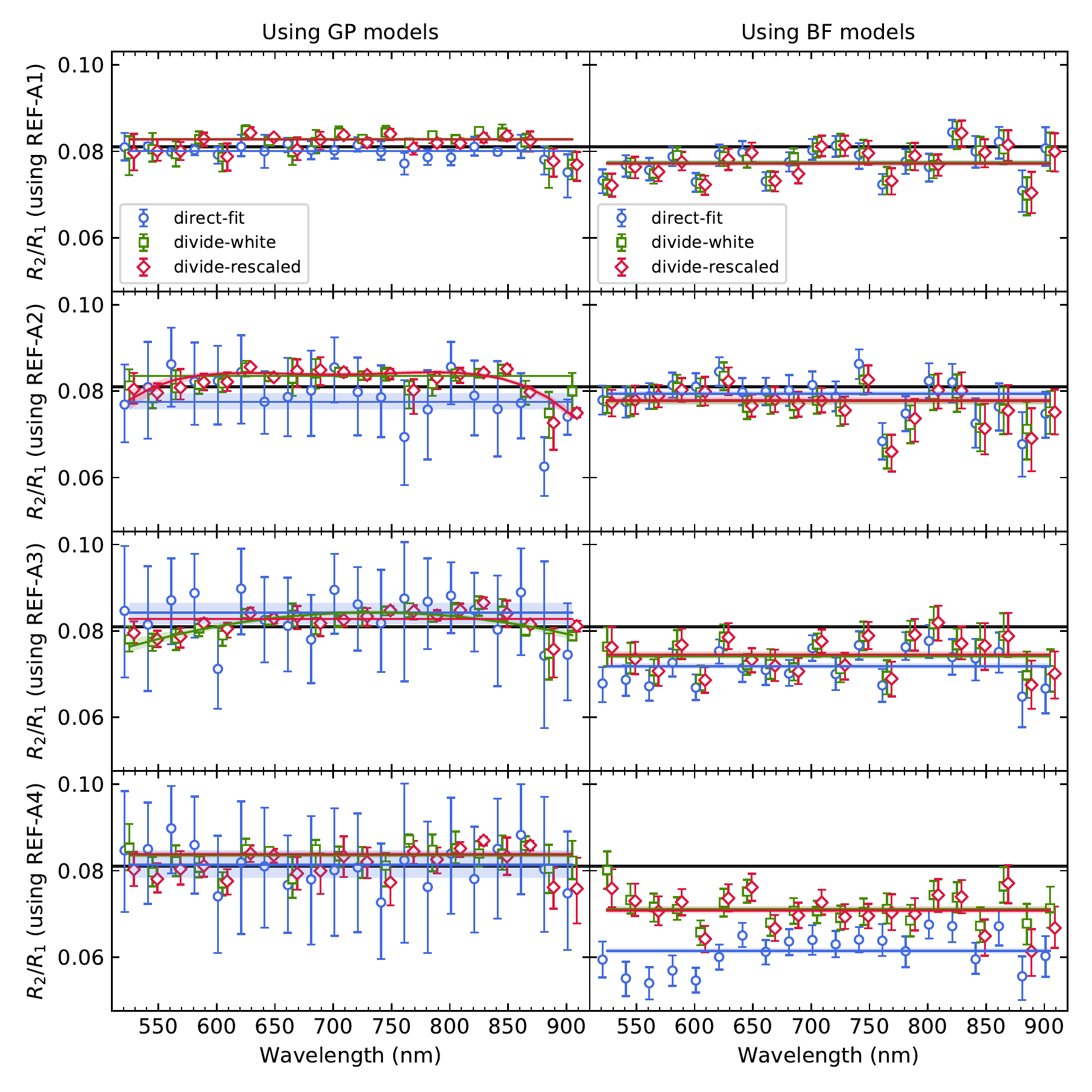}
        \caption{Transmission spectra of KIC 10657664B calculated using different noise models ({\it Left column}: GP-M32 model. {\it Right column}: Best BF model), different reference stars (REF-A1 to A4 from the top row to the bottom row), and different common-mode removal techniques (\texttt{direct-fit}: blue circles; \texttt{divide-white}: green squares; \texttt{divide-rescaled}: red diamonds). The solid lines and shaded areas are the best-fit polynomials with the strongest Bayesian evidence and their 1$\sigma$ credible intervals, respectively. The black horizontal lines are the broadband radius ratio from the {\it Kepler} data. The data points are slightly shifted along the wavelength for clarity.}
        \label{fig_compare_KIC10657664}
    \end{figure*}

    Figure \ref{fig_compare_KIC10657664} shows the transmission spectra of KIC 10657664B computed with different reference stars (REF-A1 to A4), different noise models (the GP-M32 models and the best BF models listed in Table \ref{table_bf_evidence_1065}), and different common-mode removal techniques (\texttt{direct-fit}, \texttt{divide-white}, and \texttt{divide-rescaled}). Although most of the derived spectra can be considered to be flat and featureless, only the result corresponding to REF-A1, GP, and \texttt{direct-fit} has a relatively good precision and is consistent with the broadband transit depth. The statistical metrics for the spectra in Fig. \ref{fig_compare_KIC10657664} are listed in Table \ref{table_metrics}.
 
    {\it Different reference stars.}
    The transmission spectra corresponding to REF-A1, whose light curves have weaker systematics, are closer to the broadband transit depth and have stronger evidence to be featureless than the results from the remaining reference stars. This indicates that the amplitudes of systematics in spectroscopic light curves are critical for obtaining an accurate transmission spectrum. Using different reference stars may cause considerable discrepancies in transmission spectra due to different levels and trends of light-curve systematics. We note that it is commonly impractical to simultaneously monitor multiple reference stars when using long-slit spectroscopy. A reference star should be selected before the observation based on its physical properties, including the stellar type, magnitude, and light variation. Thus by comparing the results from different reference stars, we do not strive to select a proper reference star based on the resulting transmission spectrum. Instead, we use the different reference stars to represent the inconsistent instrument-specific systematics that would exist in different observations. We show that such discrepancies in light-curve systematics can significantly impact the results.
    
    {\it Different noise models.}
    In Fig. \ref{fig_compare_KIC10657664}, the GP models and the BF models lead to totally different results, indicating that the adopted noise model also has a large impact on the transmission spectrum. When using the GP model to fit the light-curve systematics, there is no significant transit depth offset in the derived transmission spectra. When the BF models are used, however, the transit depth offsets are quite large when the light-curve systematics are too strong to be corrected by the BF model (see the results using REF-A3 and REF-A4 in Fig. \ref{fig_compare_KIC10657664}). This is mainly because in spectroscopic light-curve fitting, the orbital parameters (semimajor axis, inclination, mid-transit epoch) are fixed to the best-fit values from broadband light-curve fitting, which gives critical constraints on the transit model. However, if the adopted best-fit orbital parameters are biased from true values, the derived spectra may present considerable offsets. When GP models are used to fit the white-light curves, the derived orbital parameters, especially the mid-transit epoch, are basically consistent among four groups of data. When the BF models are used, however, the best-fit orbital parameters are less consistent. Therefore, the derived spectra using the GP models are closer to the correct transit depth than those using the BF models, as shown in Fig. \ref{fig_compare_KIC10657664}. Another reason for the results of BF models being less stable is that we adopted the best BFs based on the white-light curve analysis and used them to fit spectroscopic systematics. Since the deterministic BF models are less flexible than stochastic GP models, it is difficult for the best BF model to consistently solve the systematic noise in all narrowbands, and thus result in biased estimation on chromatic transit depths.
    
    {\it Different common-mode removal methods.}
    According to Fig. \ref{fig_compare_KIC10657664}, the common-mode removal methods in spectroscopic light-curve fitting also have large impacts on transmission spectra. When GP models are used, the spectra computed from the \texttt{direct-fit} method can have much larger errors than those from the \texttt{divide-white} and \texttt{divide-rescaled} methods. We note that the GP model usually captures the smooth variations in systematic noise. Therefore, when the common modes are not reduced, the GP model will attribute the jitter patterns induced by seeing variations to uncorrelated noise (i.e., white noise), which significantly reduces the light-curve S/N and thus overestimates the transit depth errors. Such large uncertainties will negate any possible detection from an exoplanet atmosphere. When the common-modes are reduced, however, the uncertainties of transmission spectra become so small that the results may show spurious features. When BF models are used, the common modes and the remaining components in light-curve systematics are fit by the same BF, resulting in similar uncertainties.
    As previously mentioned, the general offset of a transmission spectrum can be extremely large when the BF model fails to characterize the light-curve systematics. We find that reducing the common-mode patterns can slightly reduce such an offset, but does not solve the problem. This is because the best BF selected for the broadband systematics is too complex for the remaining spectroscopic systematics after the common-mode removal, which is likely to cause overfitted models and weaken the transit signals. In our preliminary test, when using the polynomial BF up to the first order in spectroscopic light-curve fitting, there is no significant general offset in the derived transmission spectrum, but in some wavebands, the light curve cannot be correctly fit. An ideal solution is to conduct a separate BF model selection for each narrowband light curve. We did not consider this method, however, because its computational cost is high. When we compare the \texttt{divide-white} and the \texttt{divide-rescaled} methods, their results are similar. The two methods have a risk of underestimating the transit depth errors and thus result in spurious features in the derived transmission spectra. Therefore it may be unnecessary to remove the common-mode noise when there are only small amplitudes of systematics in the light curves (e.g., when fitting the light curves calibrated with REF-A1).

    \subsubsection{KIC 9164561B}
    
\begin{figure}[htbp]
        \centering
        \includegraphics[width=\linewidth]{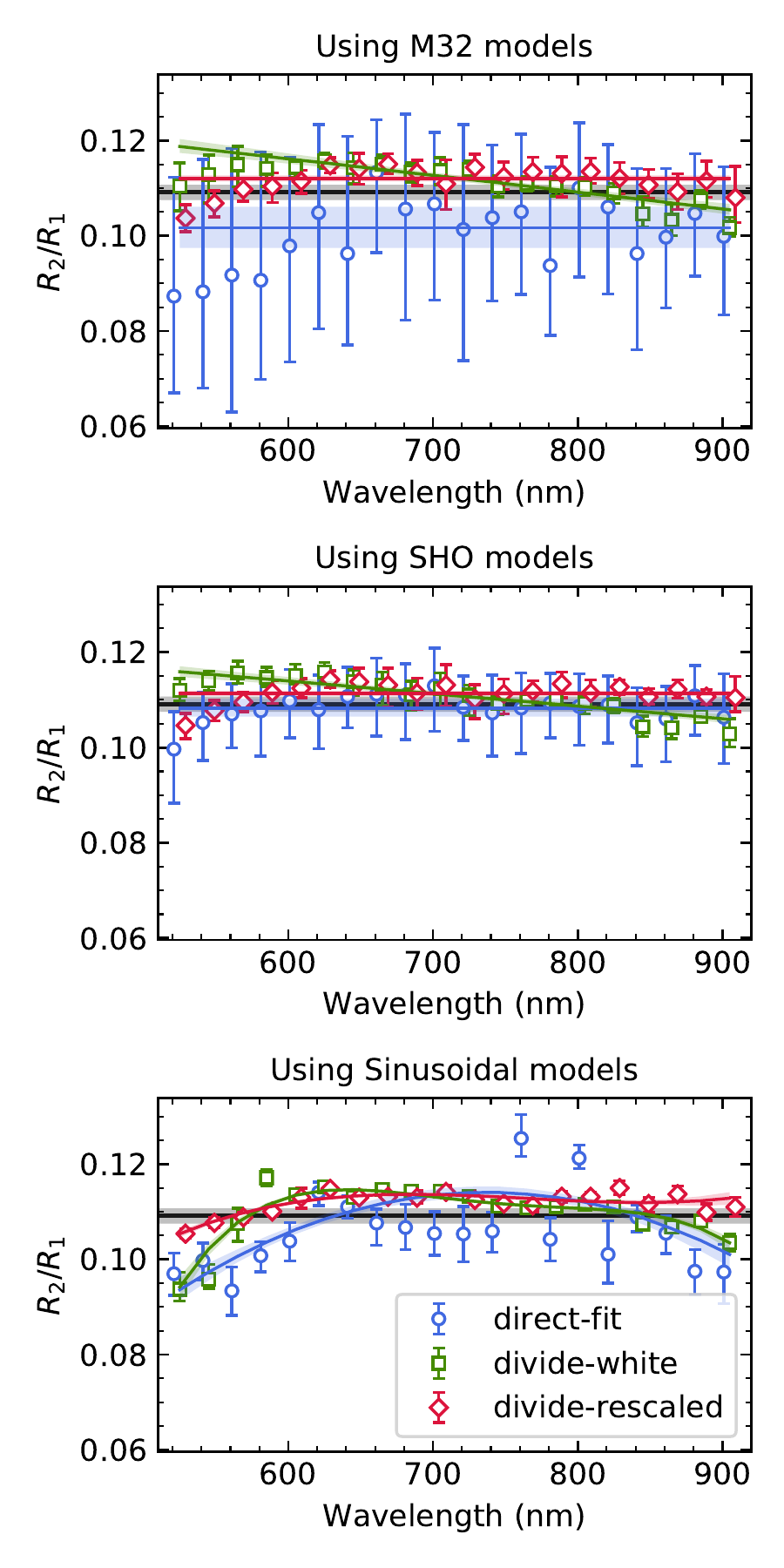}
        \caption{Transmission spectra of KIC 9164561B calculated with different noise models ({\it Top panel}: GP-M32 model. {\it Middle panel}: GP-SHO model. {\it Bottom panel}: Sinusoidal baseline model) and different common-mode removal methods (\texttt{direct-fit}: blue circles; \texttt{divide-white}: green squares; \texttt{divide-rescaled}: red diamonds). The solid lines and shaded areas are the best-fit polynomial trends with the strongest Bayesian evidence and their 1$\sigma$ credible intervals, respectively. The black lines are the broadband radius ratio from the {\it Kepler} data. The data points are slightly shifted along the wavelength for clarity.}
        \label{fig_compare_KIC9164561}
    \end{figure}

    In Fig. \ref{fig_compare_KIC9164561} we compare the transmission spectra of KIC 9164561B separately computed with two GP models (GP-M32 and GP-SHO) and one sinusoidal BF model ($m=8$), whose statistical metrics are listed in Table \ref{table_metrics}. Similar to the previous target, there are considerable discrepancies in the spectra derived from different noise models, among which the spectrum calculated with the GP-SHO model and the \texttt{divide-rescaled} method is better than other spectra considering its median uncertainty and general offsets. When the M32 kernel is used, the resulting transmission spectra have larger uncertainties, especially when the common modes are not reduced. This suggests that the M32 kernel, along with other radial basis functions, has a poor performance in extracting quasi-periodic correlated noise. When the SHO kernel is used, the derived transmission spectra have a better precision and are consistent with the broadband transit depth. In contrast, we fail to recover correct transmission spectra when the sinusoidal models are used. We speculate that this is because the sinusoidal baseline model mainly characterizes the oscillatory noise induced by stellar pulsation, but barely includes the additional patterns induced by light-curve calibration, seeing variation, and other instrumental systematics. Therefore the estimated transit depth can be significantly biased in some wavebands. Reducing the common-mode noise helps to reduce these biases, but the consequent underestimation of transit depth errors still causes spurious features in transmission spectra. The results from the \texttt{divide-rescaled} method are better than those from the \texttt{divide-white} method because the main component of systematic noise originates from stellar pulsation, which strongly depends on wavelength. The \texttt{divide-white} method would cause excess or deficient corrections on the pulsation-induced systematics and is likely to result in false linear trends when GP models are used.

    The example of KIC 9164561B indicates that the derived transmission spectra can be highly model dependent when the light-curve systematics exhibit oscillatory features and have an amplitude comparable with the transit signal. The oscillatory noise may blur the boundaries of transit ingress and egress and makes it impossible to constrain the limb-darkening features, thus impacting the estimated transit parameters. Even for flexible GP models, the results can be biased if the systematics are not characterized by suitable GPs. Therefore, it is indeed difficult to recover a transit light curve contaminated by strong stellar variations. Under these circumstances, we recommend using the SHO kernel as a robust model of stellar variations \citep{code-celerite} before we have a better understanding on the noise sources and proper modeling on their variability.
    
    \subsection{Further discussion of the flux dilution correction}
    \label{sect_flux_dilution}

    In Sect. \ref{sect_flux_dilution} we consider the flux dilution effect in transit light curves due to the strong thermal emission of hot white dwarfs. We correct for the flux dilution by calculating the flux ratios of the binaries assuming pure blackbody radiation because there is no secondary eclipse observation to directly constrain the spectroscopic flux ratio. When we use fixed effective temperatures in Eq. \ref{eq_flux_ratio}, the derived flux ratio is proportional to the transit depth. However, the relation between flux ratio and radius ratio  would be broken if the true spectroscopic flux ratio were to deviate from the blackbody hypothesis. Therefore, here we briefly discuss whether the biases of this correction would significantly impact the transmission spectra. 

    The absolute deviation of radius ratio $\delta_\mathcal{R}$ resulting from that of flux ratio $\delta_\mathcal{F}$ is 
    \begin{equation}\label{eq_error_flux_ratio}
        \delta_\mathcal{R} = \frac{\mathcal{R}_\lambda}{2(1+\mathcal{F}_\lambda)}
        \delta_\mathcal{F},
    \end{equation}
    where $\mathcal{R}_\lambda$ and $\mathcal{F}_\lambda$ denote the true values of radius ratio $R_2/R_1$ and flux ratio $F_2/F_1$ at wavelength $\lambda$. A brief derivation of Eq. \ref{eq_error_flux_ratio} is presented 
    in Appendix \ref{sect_ap_derivations}.
    
    Since the companion-to-host flux ratios for the binaries of KIC 10657664 and KIC 9164561 are measured to be $\sim$0.0189 and $\sim$0.0285 \citep{Wong2020,Rappaport2015} in the {\it Kepler} band and are far lower than 1, the relative deviation of radius ratio $\delta_\mathcal{R} / \mathcal{R}_\lambda$ then approximately equals $\delta_\mathcal{F}/2$ according to Eq. \ref{eq_error_flux_ratio}. Thus even if we did not correct the flux dilution effect and made a huge error on flux ratio that $\delta_\mathcal{F} = \mathcal{F}_\lambda \approx 0.02$, the corresponding relative deviation of radius ratio would be merely $\delta_\mathcal{R} / \mathcal{R}_\lambda \approx 1\%$. The transmission spectra in Figs. \ref{fig_compare_KIC10657664} and \ref{fig_compare_KIC9164561} have relative errors of approximately 2.5\% for KIC 10657664B and 1.8\% for KIC 9164561B when they are calculated with the GP-M32 model and the \texttt{divide-white} method. The flux dilution correction can only introduce very limited deviations that are smaller than the 1$\sigma$ credible intervals in the radius ratio. Therefore, we conclude that the significant spectral features or general offsets in the derived transmission spectra should not arise from the flux dilution effect.

    
\section{Conclusions}
\label{sect_conclusions}

    We used two transiting white dwarfs, KIC 10657664B and KIC 9164561B, as hot-Jupiter analogs to test the methods that are commonly used in low-resolution transmission spectroscopy. This benchmark test allowed us to compare the outcomes of different data analysis techniques. Its advantage is that we know beforehand that a white dwarf transmission spectrum should be flat and featureless. The transit events of these two targets were observed with the GTC OSIRIS instrument. The acquired data were reduced and analyzed in the same way as exoplanet transit spectroscopic observations. 
    
    In the analyses of the two targets, we obtained discrepant transmission spectra when we used different noise models. The results show that GP models have a better robustness than BF models when there are large amplitudes of systematic noise. However, different GP kernels may result in different transit parameters and final transmission spectra, especially when there are strong oscillatory noises in the light curves. Thus, the most suitable model must be carefully chosen based on model comparison. 
    
    In the example of KIC 10657664, the different reference stars result in different intensities of light-curve systematics that rely on specific instruments and star positions. These differences can cause significant discrepancies in the derived transmission spectra. In the particular case of GTC OSIRIS, we suggest that it is better to select a reference star closer to the target star if there are several options, given consideration of their physical properties including the spectral type, brightness, and light variations. In the example of KIC 9164561B, the GP-SHO model is found to have stronger Bayesian evidence and produces more accurate transmission spectra than the GP-M32 and the sinusoidal BF models in characterizing the stellar pulsation noise. We expect, however, that similar situations will apply to other ground-based observations of exoplanets, and the resulting large uncertainties could mute any weak exoplanet atmospheric signals.
    
    We showed that reducing the common-mode systematics with the \texttt{divide-white} or the \texttt{divide-rescaled} method helps to reduce the spectroscopic systematic noise, but also brings in the risk of underestimating the transit depth errors and introducing additional spurious spectral signatures. Thus it may be unnecessary to conduct a common-mode removal if there is only a small amplitude of systematics. 
    
    In summary, we conclude that no model exists that would always ensure a perfect characterization of the systematic noise for ground-based low-resolution transit observations. The bias of extracted systematics inevitably causes the bias of transmission spectra.  Therefore, when transmission spectroscopy studies are conducted, it is necessary to validate the results based on various model comparison, using either simulated data or real data, such as those from transiting white dwarfs, to determine their model dependence. In particular, small wiggles within the transit depth errors in a transmission spectrum should be interpreted with caution. Furthermore, the target of interest should ideally be repeatedly observed with different instruments to reduce the impact of instrumental systematics and increase the reliability of the results.

\begin{acknowledgements}
    The authors thank the anonymous referee for their valuable comments and suggestions. G.\,C. acknowledges the support by the B-type Strategic Priority Program of the Chinese Academy of Sciences (Grant No.\,XDB41000000), the National Natural Science Foundation of China (Grant No.\,42075122, 12122308), the Natural Science Foundation of Jiangsu Province (Grant No.\,BK20190110), Youth Innovation Promotion Association CAS (2021315), the China Manned Space Project (CMS-CSST-2021-B12), and the Minor Planet Foundation of the Purple Mountain Observatory. This work is based on the observations made with the Gran Telescopio Canarias installed at the Spanish Observatorio del Roque de los Muchachos of the Instituto de Astrof\'{i}sica de Canarias, in the island of La Palma.
\end{acknowledgements}

\bibliographystyle{aa}
\bibliography{reference}

\appendix
\section{Additional tables and figures}

\begin{table}[htbp]
    \centering
    \begin{threeparttable}[b]
    \caption{Prior constraints on $A_\mathrm{ELV}$ for KIC 9164561.}
    \label{table_prior_elv}
    \begin{tabular}{lccc}
        \toprule
        Waveband (nm) & 
        $u$\tnote{(1)} &
        $\tau$\tnote{(2)} &
        $A_\mathrm{ELV}$ (ppm)\tnote{(3)}  \\
        \midrule
        515 -- 535 & 0.5759 & 0.8982 & $-9995$ \\
        535 -- 555 & 0.5551 & 0.8690 & $-9745$ \\
        555 -- 575 & 0.5365 & 0.8421 & $-9521$ \\
        575 -- 595 & 0.5173 & 0.8172 & $-9308$ \\
        595 -- 615 & 0.4992 & 0.7941 & $-9113$ \\
        615 -- 635 & 0.4824 & 0.7727 & $-8934$ \\
        635 -- 655 & 0.4545 & 0.7528 & $-8721$ \\
        655 -- 675 & 0.4265 & 0.7343 & $-8520$ \\
        675 -- 695 & 0.4375 & 0.7169 & $-8477$ \\
        695 -- 715 & 0.4241 & 0.7007 & $-8346$ \\
        715 -- 735 & 0.4122 & 0.6855 & $-8227$ \\
        735 -- 755 & 0.3998 & 0.6712 & $-8111$ \\
        755 -- 775 & 0.3891 & 0.6577 & $-8007$ \\
        775 -- 795 & 0.3772 & 0.6451 & $-7904$ \\
        795 -- 815 & 0.3682 & 0.6331 & $-7815$ \\
        815 -- 835 & 0.3597 & 0.6218 & $-7732$ \\
        835 -- 855 & 0.3426 & 0.6111 & $-7623$ \\
        855 -- 875 & 0.3195 & 0.6010 & $-7498$ \\
        875 -- 895 & 0.3317 & 0.5914 & $-7493$ \\
        895 -- 915 & 0.3411 & 0.5823 & $-7482$ \\
        \bottomrule
    \end{tabular}

    {\bf Notes.} (1) Linear limb-darkening coefficients calculated with the ATLAS model of KIC 9164561 \citep{Kurucz1979,code-limb-darkening}. (2) Gravity-darkening coefficients calculated with Eq. (10) in \cite{Morris1985}. (3) Amplitudes of the ellipsoidal variation defined in Eq. \ref{eq_harmonic_correction}. The estimated uncertainties are $\sim$400 ppm according to the error propagation.

    \end{threeparttable}
\end{table}

\begin{figure}[htbp]
    \centering
    \includegraphics[width=\linewidth]{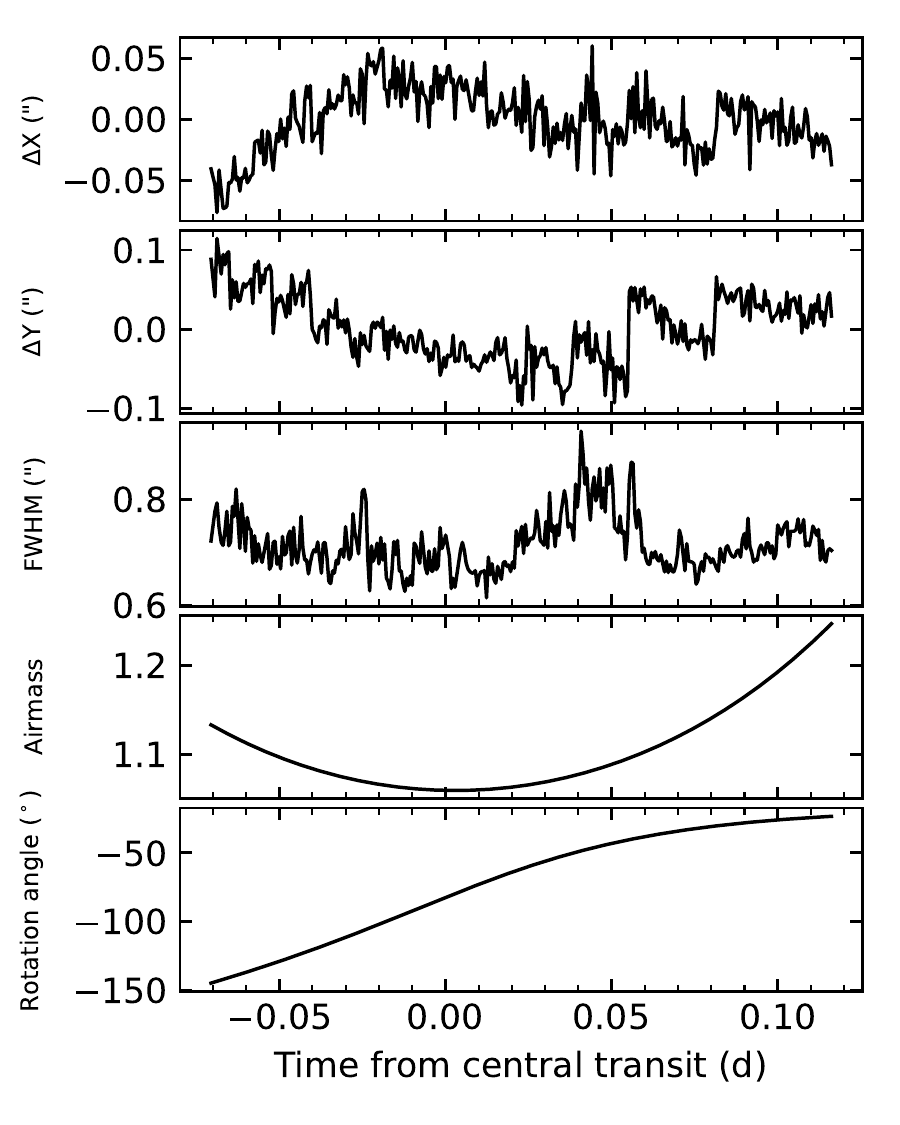}
    \caption{Time series of the state vectors used in the BF models for KIC 10657664. From top to bottom: Target position drift in the dispersion direction, target position drift in the spatial direction, the FWHM of the target PSF in the spatial direction, the airmass, and the instrumental rotation angle.}
    \label{fig_state_vectors}
\end{figure}

\begin{table*}[htbp]
    \renewcommand\arraystretch{1.2}
    \centering
    \begin{threeparttable}[b]
    \caption{Transit parameters of KIC 10657664 derived with four reference stars\tnote{(1)}.}
    \label{table_transit_parameters_1065}
    \begin{tabular}{llcccc}
        \toprule
        Parameters  & 
            Priors &
            \multicolumn{4}{c}{Posteriors} \\
         & & REF-A1 & REF-A2 & REF-A3 & REF-A4 \\
        \midrule
        $R_2/R_1$ &
            $\mathcal{N}(0.0810, 0.0001)$\tnote{(2)}  &
            $0.0810^{+0.0001}_{-0.0001}$ &
            $0.0810^{+0.0001}_{-0.0001}$ &
            $0.0810^{+0.0001}_{-0.0001}$ &
            $0.0810^{+0.0001}_{-0.0001}$ \\
        $a / R_1$ &
            $\mathcal{N}(7.05, 0.02)$\tnote{(2)} &
            $7.04^{+0.02}_{-0.02}$ &
            $7.05^{+0.02}_{-0.02}$ &
            $7.05^{+0.02}_{-0.02}$ &
            $7.05^{+0.02}_{-0.02}$ \\
        $i$ (deg)&
            $\mathcal{N}(84.52, 0.03)$\tnote{(2)} &
            $84.52^{+0.03}_{-0.03}$ & 
            $84.52^{+0.03}_{-0.03}$ & 
            $84.52^{+0.03}_{-0.03}$ & 
            $84.52^{+0.03}_{-0.03}$ \\
        $t_\mathrm{c}$ ($\rm BJD_{TBD}$)&
            $\mathcal{U}(-0.02, 0.02)$\tnote{(3)}   &
            $-0.00022^{+0.00029}_{-0.00026}$  &
            $-0.00150^{+0.00123}_{-0.00100}$  &
            $-0.00121^{+0.00105}_{-0.00099}$  &
            $-0.00133^{+0.00117}_{-0.00117}$ \\
        $u_1$ &
            $\mathcal{N}(0.24, 0.10)$\tnote{(4)} &
            $0.26^{+0.07}_{-0.07}$  &
            $0.24^{+0.09}_{-0.09}$  &
            $0.23^{+0.09}_{-0.09}$  &
            $0.23^{+0.10}_{-0.10}$  \\
        $u_2$ &
            $\mathcal{N}(0.29, 0.10)$\tnote{(4)}   &
            $0.31^{+0.08}_{-0.08}$  &  
            $0.29^{+0.09}_{-0.09}$  &
            $0.28^{+0.09}_{-0.09}$  &
            $0.28^{+0.10}_{-0.10}$\\
        \bottomrule
    \end{tabular}
    
    {\bf Notes.} (1) Using the GP-M32 model in light-curve fitting. (2) Constrained by the estimates of \cite{Wong2020}. (3) Subtracted by 2457220.532790 based on the transit ephemeris of \cite{Wong2020}. (4) Constrained by the ATLAS stellar atmosphere model of the host star.

    \end{threeparttable}
\end{table*}

\begin{table*}[htbp]
    \renewcommand\arraystretch{1.2}
    \centering
    \begin{threeparttable}[b]
    \caption{Transit parameters of KIC 9164561.}
    \label{table_transit_parameters_9164}
    \begin{tabular}{llccc}
        \toprule
        Parameters  & 
            Priors &
            \multicolumn{3}{c}{Posteriors} \\
         & & M32 & SHO & Sinusoidal \\
        \midrule
        $R_2/R_1$ &
            $\mathcal{N}(0.1091, 0.0016)$\tnote{(1)}  &
            $0.1090^{+0.0015}_{-0.0015}$ &
            $0.1088^{+0.0014}_{-0.0016}$ &
            $0.1091^{+0.0014}_{-0.0013}$\\
        $a / R_1$ &
            $\mathcal{N}(2.52, 0.04)$\tnote{(1)}  &
            $2.55^{+0.03}_{-0.04}$ &
            $2.55^{+0.03}_{-0.03}$ &
            $2.53^{+0.02}_{-0.02}$ \\
        $i$ (deg)&
            $\mathcal{N}(71.59, 0.22)$\tnote{(1)}     &
            $71.53^{+0.19}_{-0.20}$ &
            $71.51^{+0.20}_{-0.20}$ &
            $71.46^{+0.17}_{-0.17}$\\
        $t_\mathrm{c}$  ($\rm BJD_{TBD}$)&
            $\mathcal{U}(-0.02, 0.02)$\tnote{(2)}   &
            $-0.00433^{+0.00476}_{-0.00442}$  &
            $-0.01026^{+0.00443}_{-0.00334}$  &
            $-0.00826^{+0.00058}_{-0.00051}$  \\
        $u_1$ &
            $\mathcal{N}(0.21, 0.10)$\tnote{(3)}   &
            $0.23^{+0.09}_{-0.10}$ &
            $0.25^{+0.09}_{-0.09}$ &
            $0.26^{+0.07}_{-0.07}$ \\
        $u_2$ &
            $\mathcal{N}(0.33, 0.10)$\tnote{(3)}   &
            $0.34^{+0.10}_{-0.09}$ &
            $0.35^{+0.09}_{-0.09}$ & 
            $0.35^{+0.09}_{-0.08}$\\
        \bottomrule
    \end{tabular}

    {\bf Notes.} (1) Constrained by the estimates of \cite{Rappaport2015}. (2) Subtracted by 2457213.61556 based on the transit ephemeris of \cite{Rappaport2015}. (3) Constrained by the ATLAS model of the host star.

    \end{threeparttable}
\end{table*}

\begin{table*}[htbp]
    \centering
    \begin{threeparttable}[b]
    \caption[]{Top three BFs from model selection
                in the white-light curve analyses of KIC 10657664.}
    \label{table_bf_evidence_1065}
    \begin{tabular}{llcc}
        \toprule
        \# \tnote{(1)}         &
        BF \tnote{(2)}         & 
        $D$ \tnote{(3)}        &
        $\ln\mathcal{Z}$ \tnote{(4)} \\
        \midrule
        \multicolumn{3}{c}{REF-A1} \\
        1 &
        $c_0 + \sum_{n=1}^2 c_{a,n} a^n + \sum_{n=1}^2 c_{w,n} w^n$ &
        12 &
        $2237.97 \pm 0.20$ \\
        2 &
        $c_0 + \sum_{n=1}^2 c_{a,n} a^n + \sum_{n=1}^1 c_{w,n} w^n + \sum_{n=1}^1 c_{x,n} x^n$ &
        12 &
        $2237.29 \pm 0.21$ \\
        3 &
        $c_0 + \sum_{n=1}^3 c_{a,n} a^n + \sum_{n=1}^1 c_{w,n} w^n$ &
        12 &
        $2237.05 \pm 0.20$ \\

        \multicolumn{3}{c}{REF-A2} \\
        1 &
        $c_0 + \sum_{n=1}^3 c_{a,n} a^n + \sum_{n=1}^1 c_{w,n} w^n + \sum_{n=1}^3 c_{z,n} z^n$ &
        15 &
        $2140.37 \pm 0.23$ \\
        2 &
        $c_0 + \sum_{n=1}^3 c_{a,n} a^n + \sum_{n=1}^1 c_{w,n} w^n + \sum_{n=1}^1 c_{x,n} x^n + \sum_{n=1}^3 c_{z,n} z^n$ &
        16 &
        $2136.23 \pm 0.24$ \\
        3 &
        $c_0 + \sum_{n=1}^3 c_{a,n} a^n + \sum_{n=1}^2 c_{w,n} w^n + \sum_{n=1}^3 c_{z,n} z^n$ &
        16 &
        $2135.72 \pm 0.24$ \\
        
        \multicolumn{3}{c}{REF-A3} \\
        1 &
        $c_0 + \sum_{n=1}^3 c_{a,n} a^n + \sum_{n=1}^1 c_{w,n} w^n + \sum_{n=1}^1 c_{y,n} y^n + \sum_{n=1}^3 c_{z,n} z^n$ &
        16 &
        $2057.02 \pm 0.24$ \\
        2 &
        $c_0 + \sum_{n=1}^3 c_{a,n} a^n + \sum_{n=1}^3 c_{w,n} w^n + \sum_{n=1}^1 c_{y,n} y^n + \sum_{n=1}^3 c_{z,n} z^n$ &
        18 &
        $2053.77 \pm 0.25$ \\
        3 &
        $c_0 + \sum_{n=1}^3 c_{a,n} a^n + \sum_{n=1}^1 c_{w,n} w^n + \sum_{n=1}^2 c_{y,n} y^n + \sum_{n=1}^3 c_{z,n} z^n$ &
        17 &
        $2053.38 \pm 0.25$ \\
        
        \multicolumn{3}{c}{REF-A4} \\
        1 &
        $c_0 + \sum_{n=1}^3 c_{a,n} a^n + \sum_{n=1}^2 c_{w,n} w^n + \sum_{n=1}^1 c_{y,n} y^n + \sum_{n=1}^3 c_{z,n} z^n$ &
        17 &
        $1939.96 \pm 0.24$ \\
        2 &
        $c_0 + \sum_{n=1}^3 c_{a,n} a^n + \sum_{n=1}^2 c_{w,n} w^n + \sum_{n=1}^3 c_{z,n} z^n$ &
        16 &
        $1938.71 \pm 0.23$ \\
        3 &
        $c_0 + \sum_{n=1}^3 c_{a,n} a^n + \sum_{n=1}^3 c_{w,n} w^n + \sum_{n=1}^1 c_{y,n} y^n + \sum_{n=1}^3 c_{z,n} z^n$ &
        18 &
        $1937.55 \pm 0.25$ \\
        
        \bottomrule
        \multicolumn{4}{c}{GP model with the M32 kernel} \\
        \midrule
         & REF-A1 & 9 & $2236.26 \pm 0.10$ \\
         & REF-A2 & 9 & $2030.73 \pm 0.10$ \\
         & REF-A3 & 9 & $2094.07 \pm 0.10$ \\
         & REF-A4 & 9 & $1952.73 \pm 0.10$ \\
        \bottomrule
    \end{tabular}
    
    {\bf Notes.} (1) Order of BFs sorted by $\ln\mathcal{Z}$. (2) $c$, baseline coefficients; $a$, rotation angle of the instrument; $w$, FWHM of stellar PSF along the spatial direction;  $x$, stellar position drift along the dispersion direction; $y$, stellar position drift along the spatial direction; $z$, airmass. (3) Total model dimensionality including six transit parameters and one jitter term. (4) Bayesian model evidence.

    \end{threeparttable}
\end{table*}    

\begin{table*}[htbp]
    \centering
    \begin{threeparttable}[b]
    \caption[]{Bayesian evidence of the noise models 
        in the white-light curve analyses of KIC 9164561.}
    \label{table_bf_evidence_9164}
    \begin{tabular}{llc}
        \toprule
        \multicolumn{3}{c}{Sinusoidal BF models} \\
        Number of sinusoidal terms\tnote{(1)} & 
        $D$\tnote{(2)} &
        $\ln\mathcal{Z}$\tnote{(3)} \\
        \midrule
        $m=1$ &
        13 &
        $704.08 \pm 0.24$ \\
        $m=2$ &
        16 &
        $765.50 \pm 0.30$ \\
        $m=3$ &
        19 &
        $867.09 \pm 0.39$ \\
        $m=4$ &
        22 &
        $879.79 \pm 0.40$ \\
        $m=5$ &
        25 &
        $883.48 \pm 0.42$ \\
        $m=6$ &
        28 &
        $892.99 \pm 0.40$ \\
        $m=7$ &
        31 &
        $896.02 \pm 0.39$ \\
        $m=8$ &
        34 &
        $900.25 \pm 0.43$ \\
        \bottomrule
        \multicolumn{3}{c}{GP models} \\
        kernel &
        $D$ &
        $\ln\mathcal{Z}$ \\
        \midrule
        M32 &
        11 &
        $914.51 \pm 0.09$ \\
        SHO &
        12 &
        $916.20 \pm 0.10$ \\
        \bottomrule
    \end{tabular}
    
    {\bf Notes.} (1) Defined in Eq. \ref{eq_sinusoidal_bf}. (2) Total model dimensionality including six transit parameters, two baseline variation parameters, and a jitter term. (3) Bayesian model evidence.

    \end{threeparttable}
\end{table*}  

\begin{figure*}[htbp]
    \centering
    \includegraphics[width=\linewidth]{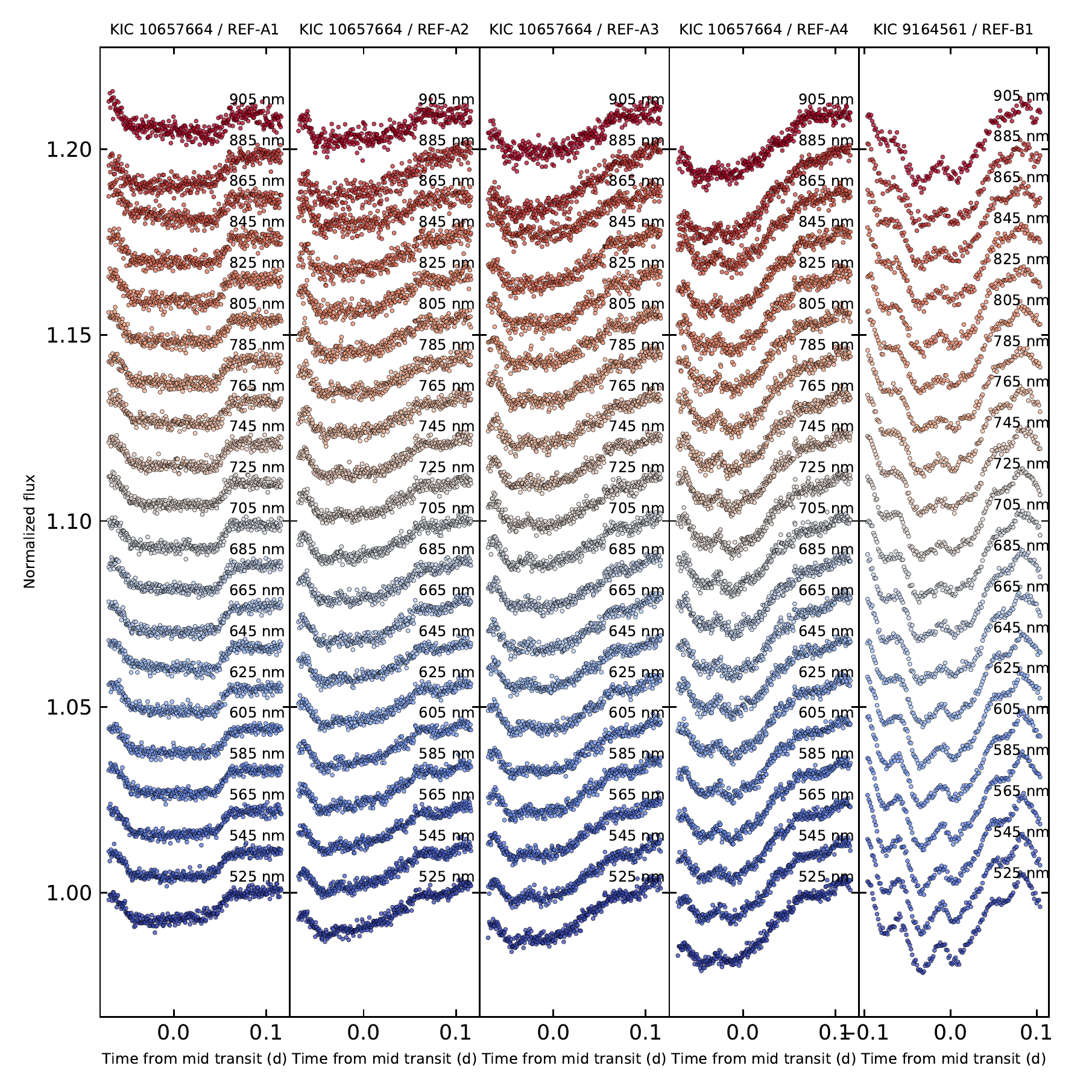}
    \caption{Raw spectroscopic light curves of KIC 10657664 and KIC 9164561.
            The colors of the light curves indicate corresponding wavebands.
            The light curves are arbitrarily shifted for clarity.
            The flux errors are too small to be shown.}
    \label{fig_spec_lightcurves}
\end{figure*}

\begin{table*}[htbp]
    \small
    \centering
    \begin{threeparttable}[b]
    \caption{Statistical metrics for the transmission spectra of KIC 10657664B and KIC 9164561B.}
    \label{table_metrics}
    \begin{tabular}{lllcccccc}
        \toprule
        \multicolumn{9}{c}{KIC 10657664B} \\
        \toprule
        \makecell{Reference \\ star} & 
        \makecell{Noise \\ model} & 
        \makecell{Common-mode \\ removal technique} & 
        \makecell{Median \\ RMSE/$\varepsilon_{\rm ph}$}\tnote{(1)}&
        \makecell{Median \\ $\varepsilon_{R_2/R_1}$}\tnote{(2)} &
        Mean $R_2/R_1$\tnote{(3)} & 
        \makecell{$t$-score ($\sigma$) \\of the offset}\tnote{(4)} &  
        Feature\tnote{(5)} & 
        \makecell{$\chi^2_\nu$ of the\\feature}\tnote{(6)} \\
        \midrule
        REF-A1 & 
            GP-M32 & 
                \texttt{direct-fit} & 
                1.09 & 0.0021 & $0.08008 \pm 0.00030$ & 2.9 & Flat & 0.36\\ 
                & & \texttt{divide-white} & 
                1.06 & 0.0018 & $0.08291 \pm 0.00024$ & 7.2 & Flat & 1.12\\
                & & \texttt{divide-rescaled} & 
                1.06 & 0.0015 & $0.08264 \pm 0.00030$ & 5.2 & Flat & 0.87\\
                \cmidrule(l){2-9}
            & BF \#1 &
                \texttt{direct-fit} & 
                1.10 & 0.0026 & $0.07729 \pm 0.00060$ & 6.1 & Flat & 1.57\\
                & & \texttt{divide-white} & 
                1.06 & 0.0024 & $0.07727 \pm 0.00057$ & 6.5 & Flat & 1.69\\
                & & \texttt{divide-rescaled} & 
                1.06 & 0.0025 & $0.07712 \pm 0.00058$ & 6.6 & Flat & 1.79\\
        \midrule
        REF-A2 & 
            GP-M32 & 
                \texttt{direct-fit} & 
                1.32 & 0.0090 & $0.07777 \pm 0.00180$ & 1.8 & Flat & 0.54\\
                & & \texttt{divide-white} & 
                1.13 & 0.0021 & $0.08362 \pm 0.00028$ & 8.9 & Flat & 0.91\\
                & & \texttt{divide-rescaled} & 
                1.13 & 0.0016 & $0.08333 \pm 0.00024$ & 9.0 & Nonlinear & 1.00\\
                \cmidrule(l){2-9}
            & BF \#1 &
                \texttt{direct-fit} & 
                1.34 & 0.0035 & $0.07950 \pm 0.00082$ & 1.8 & Flat & 1.16\\
                & & \texttt{divide-white} & 
                1.28 & 0.0034 & $0.07754 \pm 0.00078$ & 4.4 & Flat & 1.19\\
                & & \texttt{divide-rescaled} & 
                1.33 & 0.0032 & $0.07782 \pm 0.00076$ & 4.1 & Flat & 1.01\\
        \midrule
        REF-A3 & 
            GP-M32 & 
                \texttt{direct-fit} & 
                1.17 & 0.0105 & $0.08389 \pm 0.00241$ & 1.2 & Flat & 0.26\\
                & & \texttt{divide-white} & 
                1.07 & 0.0019 & $0.08239 \pm 0.00034$ & 3.9 & Nonlinear & 0.73\\
                & & \texttt{divide-rescaled} & 
                1.07 & 0.0016 & $0.08280 \pm 0.00033$ & 5.3 & Flat & 1.49\\
                \cmidrule(l){2-9}
            & BF \#1 &
                \texttt{direct-fit} & 
                1.25 & 0.0035 & $0.07188 \pm 0.00079$ & 11.4 & Flat & 1.11\\
                & & \texttt{divide-white} & 
                1.18 & 0.0035 & $0.07421 \pm 0.00076$ & 8.8 & Flat & 1.22\\
                & & \texttt{divide-rescaled} & 
                1.19 & 0.0037 & $0.07449 \pm 0.00080$ & 8.1 & Flat & 1.28\\
        \midrule
        REF-A4 & 
            GP-M32 & 
                \texttt{direct-fit} & 
                1.41 & 0.0137 & $0.08144 \pm 0.00297$ & 0.1 & Flat & 0.14\\
                & & \texttt{divide-white} & 
                1.12 & 0.0035 & $0.08396 \pm 0.00057$ & 5.1 & Flat & 0.82\\
                & & \texttt{divide-rescaled} & 
                1.11 & 0.0035 & $0.08422 \pm 0.00050$ & 6.3 & Flat & 1.32\\
                \cmidrule(l){2-9}
            & BF \#1 &
                \texttt{direct-fit} & 
                1.48 & 0.0035 & $0.06140 \pm 0.00076$ & 25.5 & Flat & 1.45\\
                & & \texttt{divide-white} & 
                1.20 & 0.0032 & $0.07124 \pm 0.00073$ & 13.3 & Flat & 0.91\\
                & & \texttt{divide-rescaled} & 
                1.21 & 0.0035 & $0.07071 \pm 0.00077$ & 13.3 & Flat & 1.22\\
        \toprule
        \multicolumn{9}{c}{KIC 9164561B} \\
        \toprule
        REF-B1 & 
            GP-M32 & 
                \texttt{direct-fit} & 
                1.18 & 0.0187 & $0.10197 \pm 0.00404$ & 1.6 & Flat & 0.14\\
                & & \texttt{divide-white} & 
                1.09 & 0.0022 & $0.11125 \pm 0.00049$ & 1.3 & Linear & 0.59\\
                & & \texttt{divide-rescaled} & 
                1.19 & 0.0029 & $0.11204 \pm 0.00063$ & 1.7 & Flat & 0.95\\
                \cmidrule(l){2-9}
            & GP-SHO & 
                \texttt{direct-fit} & 
                1.20 & 0.0077 & $0.10833 \pm 0.00171$ & 0.3 & Flat & 0.08\\
                & & \texttt{divide-white} & 
                1.12 & 0.0022 & $0.11012 \pm 0.00044$ & 0.6 & Linear & 0.46\\
                & & \texttt{divide-rescaled} & 
                1.15 & 0.0025 & $0.11137 \pm 0.00051$ & 1.4 & Flat & 0.70\\
                \cmidrule(l){2-9}
            & BF-SIN &
                \texttt{direct-fit} & 
                1.43 & 0.0044 & $0.10824 \pm 0.00083$ & 0.5 & Nonlinear & 3.11\\
                & & \texttt{divide-white} & 
                1.21 & 0.0010 & $0.11228 \pm 0.00019$ & 2.0 & Nonlinear & 2.84\\
                & & \texttt{divide-rescaled} & 
                1.21 & 0.0012 & $0.11192 \pm 0.00023$ & 1.7 & Nonlinear & 1.52\\
        \bottomrule
    \end{tabular}
    
    {\bf Notes.} (1) Median ratio of the RMSE and the expected amplitude of photon-dominated noise in spectroscopic light-curve fitting. (2) Median uncertainty of the transmission spectrum. (3) Weighted mean of the derived transmission spectrum. (4) $t$-score for the null hypothesis that the averaged spectrum is consistent with the white-light radius ratio. (5) ``Flat'' if the best-fit polynomial of the derived spectrum is zeroth-order; ``Linear'' if it is first order; ``Nonlinear'' if it is second or higher order. (6) Reduced $\chi^2$ of the best-fit polynomial for the derived spectrum.
    \end{threeparttable}
\end{table*}  

\section{Derivations of Eq. \ref{eq_error_flux_ratio}}\label{sect_ap_derivations}

    We denote the true values of the companion-to-host radius ratio and flux ratio as $\mathcal{R}$ and $\mathcal{F}$, respectively. The flux dilution correction in Eq. \ref{eq_dilution} is equivalent to 
    \begin{equation} \label{eq_dilution_2}
        1 - m^*(t) = \frac{1 - m(t)}{1 + \mathcal{F}},
    \end{equation}
    where $m^*(t)$ is the diluted transit model, while $m(t)$ is the model without the dilution effect.
    
    At the mid-transit epoch $t_\mathrm{c}$, there is a linear approximation between the normalized flux and the transit depth 
    \begin{equation} \label{eq_transit_depth}
    \begin{split}
        m|_{t=t_\mathrm{c}} &=  1 - \xi \mathcal{R}^2 + o(\mathcal{R}^2), \\
        m^*|_{t=t_\mathrm{c}} &= 1 - \xi \mathcal{R}^{*2} + o(\mathcal{R}^{*2}),        
    \end{split}
    \end{equation}
    where $\xi$ is a constant accounting for the host star's limb-darkening effect, and $\mathcal{R^*}$ is the diluted radius ratio. Substituting Eqs. \ref{eq_transit_depth} into Eq. \ref{eq_dilution_2}, we eliminate $\xi$ and obtain
    \begin{equation} \label{eq_RF1}
        \mathcal{R} = \sqrt{1+\mathcal{F}} \mathcal{R}^*.
    \end{equation}
    
    When error propagation is considered, the deviation of radius ratio $\delta_\mathcal{R}$ arising from that of flux ratio $\delta_\mathcal{F}$ writes 
    \begin{equation} \label{eq_RF2}
        \delta_\mathcal{R} = \frac{\partial \mathcal{R}}{\partial\mathcal{F}}
        \delta_\mathcal{F} = \frac{\mathcal{R}}{2(1+\mathcal{F})} \delta_\mathcal{F},
    \end{equation}    
    where $\mathcal{R}^*$ is eliminated. Thus for $0 < \mathcal{F} \ll 1$, the relative deviation of radius ratio is
    \begin{equation}
        \frac{\delta_\mathcal{R}}{\mathcal{R}} \approx 0.5\delta_\mathcal{F}.
    \end{equation}    
    
\end{document}